\documentclass{UApreprint}
\usepackage{epsfig}
\usepackage{graphics}
\usepackage{fullpage}
\usepackage{verbatim}
\usepackage[tight]{subfigure}
\usepackage{amssymb}
\newcommand{\be}{\begin{equation}}
\newcommand{\ee}{\end{equation}}
\newcommand{\bea}{\begin{eqnarray}}
\newcommand{\eea}{\end{eqnarray}}
\newcommand{\bean}{\begin{eqnarray*}}
\newcommand{\eean}{\end{eqnarray*}}
\newcommand{\bi}{\begin{itemize}}
\newcommand{\ei}{\end{itemize}}
\newcommand{\bdm}{\begin{displaymath}}
\newcommand{\edm}{\end{displaymath}}

\newcommand{\unit}{\mathbf{I}}
\newcommand{\tr}{\mathrm{Tr}}
\newcommand{\Ye}{\mathbf{Y}_e}
\newcommand{\Yn}{\mathbf{Y}_\nu}
\newcommand{\Yd}{\mathbf{Y}_d}
\newcommand{\Yu}{\mathbf{Y}_u}
\newcommand{\Ae}{\mathbf{A}_e}
\newcommand{\An}{\mathbf{A}_\nu}
\newcommand{\Ad}{\mathbf{A}_d}
\newcommand{\Au}{\mathbf{A}_u}
\newcommand{\mls}{\mathbf{m}^2_{\tilde{L}}}
\newcommand{\mes}{\mathbf{m}^2_{\tilde{e^c}}}
\newcommand{\mns}{\mathbf{m}^2_{\tilde{\nu}}}
\newcommand{\mqs}{\mathbf{m}^2_{\tilde{q}}}
\newcommand{\mds}{\mathbf{m}^2_{\tilde{d^c}}}
\newcommand{\mus}{\mathbf{m}^2_{\tilde{u^c}}}
\newcommand{\mhds}{m^2_{H_d}}
\newcommand{\mhus}{m^2_{H_u}}

\newcommand{\twi}{\widetilde}
\newcommand{\smass}[1]{m^2_{\twi #1}}

\begin{document}
%\draft

\title{Observing The Hidden Sector}

\author{Bruce A. Campbell~$^{1,2}$, John Ellis~$^2$ and David W. Maybury~$^1$}
\address{$^1$Department of Physics, Carleton University,
1125 Colonel By Drive, \\ Ottawa ON K1S 5B6, Canada \\
$^2$ Theory Division, PH Department,
CERN, CH-1211 Geneva 23, Switzerland}

\date{October 27th, 2008}

\abstract{We study the effects of renormalization due to hidden-sector dynamics on 
observable soft supersymmetry-breaking parameters in the minimal supersymmetric
extension of the Standard Model (MSSM), under various hypotheses about their
universality at a high input scale. We show that hidden-sector renormalization effects
may induce the spurious appearance of unification of the scalar masses
at some lower scale, as in mirage unification scenarios. We demonstrate in simple
two-parameter models of the hidden-sector dynamics that the parameters may in
principle be extracted from experimental measurements, rendering the hidden
sector observable. We also discuss the ingredients that would be necessary to carry this
programme out in practice.}

\archive{}
\preprintone{CERN-PH-TH/2008-209}
\preprinttwo{}

\submit{}

\maketitle

%%%%%%%%%%%%%%%%%%%%%%%%%%%%%%%%%%%%%%%%%%%%%%%%%%%%%%%
%%%%%%%%%%%%%%%%%%%%%%%%%%%%%%%%%%%%%%%%%%%%%%%%%%%%%%%
%%%%%%%%%%%%%%%%%%%%%%%%%%%%%%%%%%%%%%%%%%%%%%%%%%%%%%%
%%%%%%%%%%%%%%%%%%%%%%%%%%%%%%%%%%%%%%%%%%%%%%%%%%%%%%%

\section{Introduction}

The most problematic aspect of supersymmetric phenomenology is the mechanism whereby supersymmetry is broken. The scenario usually adopted is that supersymmetry is broken in some `hidden' sector that is (almost) decoupled from the observable 
sector~\cite{Polonyi,Many1,Many2,Nilles,DIN}. 
It is often suggested, primarily for reasons of simplicity, that supersymmetry breaking is universal at some high input renormalization scale, perhaps the scale of grand or string unification~\cite{SUSY07}.
This scenario, known as the
CMSSM, is certainly simple, but it is not necessarily favoured by specific models of the dynamics in the hidden sector.

When relating this or any other scenario for supersymmetry breaking to low-energy phenomenology, it has usually been assumed that the renormalization of the effective soft supersymmetry-breaking parameters of the MSSM may be calculated
reliably using the renormalization-group equations (RGEs) of the MSSM, neglecting the dynamics of the hidden sector. However, it was recently pointed out~\cite{din:04,crs:07,mnp:07} that this may not be the case, and that the renormalization of the observable-sector supersymmetry-breaking parameters may be sensitive also to the hidden-sector dynamics~\cite{kkm:08}. 
The bad news is that this sensitivity introduces additional ambiguity into the low-energy predictions of even the CMSSM:  the good news is that this sensitivity may enable experiment to provide some insight into the mechanism of supersymmetry breaking, even if it is `hidden', and thought previously to be inaccessible to experimental probes. Thus, even the hidden sector may be observable.

We address this possibility in the context of supersymmetric theories with gauge coupling unification at a unification scale $M_{X}$. For simplicity, we further assume that supersymmetry breaking is mediated to the observable sector at a (high) scale $M$ that is at least as large as the unification scale (this is the case in gravitational/moduli mediation and in anomaly mediation, where the scale is typically of order the reduced Planck mass). The latter assumption is not essential to our argument, and variants of our proposal can be adapted to lower-scale mediation mechanisms such as gauge mediation.

We show in this paper how, in a simple but general parametrization of the hidden sector model,
low-energy measurements may be used, in principle, to extract information about the hidden sector.
We work out in detail 
in this paper the reconstruction of the hidden sector in models with the following
two parameters: the dynamical scale $M_{Hid}$ of the hidden sector, which is the infrared cutoff of the hidden sector running, and its effective interaction strength 
$\lambda$ at the mediation scale. This particular parametrization has already been used 
previously in studies of hidden-sector effects~\cite{din:04,crs:07}. Further examples of calculations 
of the effects of 
the hidden sector in models parametrized in this way appear in a companion paper~\cite{cem:2},
which extends the discussion of this paper to seesaw models of neutrino masses and 
flavour mixing.

We adopt a renormalization scheme in which the hidden-sector 
external-line wave-function renormalization of the singlet superfield $S$ (whose F-term VEV is responsible for supersymmetry breaking) is absorbed into that VEV.
The RGE evolutions of the gaugino masses are then unaffected by the hidden sector. 
However, the RGE evolutions of the supersymmetry-breaking scalar mass parameters 
also pick up one-particle-irreducible (1PI) contributions, and are in general modified by 
hidden-sector running down to the hidden-sector scale $M_{Hid}$, by an amount that depends on 
the effective interaction strength $\lambda$. As a result, the apparent scalar-mass unification scale 
that would, in our example, normally be inferred in the CMSSM from low-energy measurements is in general decreased below the true unification scale determined by the gauge couplings, approaching $M_{Hid}$ in the limit of large $\lambda$. More generally, depending on the sign of the hidden-sector effects, the unification could also be `blown up'. The resulting distortion of the scalar spectrum is a characteristic of the hidden sector, which may be inverted to characterize general two-parameter 
models of the hidden-sector scalar-mass operator renormalization, giving information on the 
general structure of the hidden sector. This may be used to fit the two parameters of the hidden
sector, and thereby render it `observable', in a limited and indirect sense. 

We give numerical examples of the modified RGE effects on different supersymmetry-breaking scalar mass parameters, and illustrate explicitly how low-energy measurements may be used to determine $M_{Hid}$ and $\lambda$. Since there are many such mass parameters, namely 
$m_{L, E^c, Q, U^c, D^c}$ for each of the three generations of sleptons and squarks, 
there is considerable redundancy in the determination of $M_{Hid}$ and $\lambda$, 
and the universality assumed in the CMSSM can be tested in parallel.

In fact, as we discuss below, even if one assumes only that scalars with the same gauge 
charges have the same mass at the unification scale (as is strongly indicated by the stringent 
limits on flavour-violating neutral interactions~\cite{EN,BG,il:02,cam:02}) then the interference of 
hidden-sector renormalization effects with observable-sector effects induced by the top-quark 
Yukawa coupling will allow one
to fit the two parameters of the models of the hidden sector that we consider. If 
$\tan\beta$ is sufficiently large for effects of the bottom Yukawa coupling on the RGE flow to
be distinguishable, then an independent determination of the parameters in a two-parameter hidden sector model is possible. In general, comparison of these determinations will yield the same parameter 
values only if the parametrization in which the determination is done is one that correctly describes 
the qualitative behaviour of the hidden sector. If this is achieved, we would be able not only to select 
the correct hidden-sector parametrization, but also to determine the numerical values of its parameters.

In string constructions with gauge-coupling unification, there are often
characteristic patterns of both the gaugino and scalar soft mass terms. For example, in 
F-theory constructions of unified theories with moduli mediation of flux breaking of supersymmetry, 
it has recently been argued that there is a simple pattern of soft gaugino terms, of soft scalar 
mass terms, and of A-terms, of the standard modulus-dominated type~\cite{iba:08}. Similarly,
in heterotic constructions of unified theories with uplifting via matter superpotentials, one typically 
sees a  characteristic `mirage' pattern for the soft terms, with the relative strength of the moduli contributions to the anomaly contributions governed by a single parameter $\rho$~\cite{nil:08}. In 
different four-dimensional supersymmetric GUT models, various sets of relations between
the input soft parameters are possible: for example, in supersymmetric SO(10) all the soft
superysmmetric scalar masses would be equal, whereas in conventional SU(5) the masses of
scalars in the $\mathbf{\bar 5}$ and $\mathbf{10}$ representations would be different in
general. In cases like these there are many relations 
between the soft parameters at the unification scale, which mean that there are many redundant 
checks on the choice and parameter values of a hidden-sector model, which can be 
fitted using the observed values of the soft parameters at low energy.

The organization of our paper is as follows. In Section 2 we review the detailed nature of the 
possible hidden-sector renormalization effects. In Section 3 we introduce simple generic families of 
two-parameter hidden-sector models parametrized by the nature of the induced renormalization of the 
scalar mass operators. In Section 4 we demonstrate explicitly our proposed reconstruction in the case 
of a toy two-parameter hidden-sector model of the kind that was discussed in Section 3, but using 
a different parametrization taken from the existing literature, in order to facilitate the comparison of 
our results to those already extant. In Section 5 we discuss how one would undertake the 
proposed reconstruction starting with experimental data from the LHC and a future linear
collider. Section 6 presents our conclusions. The Appendices list scalar-mass sum rules and
the relevant RGEs.

%%%%%%%%%%%%%%%%%%%%%%%%%%%%%%%%%%%%%%%%%%%%%%%%%%%%%%%
%%%%%%%%%%%%%%%%%%%%%%%%%%%%%%%%%%%%%%%%%%%%%%%%%%%%%%%
%%%%%%%%%%%%%%%%%%%%%%%%%%%%%%%%%%%%%%%%%%%%%%%%%%%%%%%
%%%%%%%%%%%%%%%%%%%%%%%%%%%%%%%%%%%%%%%%%%%%%%%%%%%%%%%

\section{Hidden-Sector Renormalization}
\label{sec:mech}

In this section, we analyze the renormalization 
of operators that couple the hidden and CMSSM fields, 
as would result from 
strongly-coupled dynamics in the hidden sector~\cite{din:04,mnp:07}.

In general, there are both gauge 
non-singlet and singlet fields in the hidden sector,  which we 
call $F$ and $S$, respectively, without referring to particular 
models; they may be ``elementary'' or ``composite''. 
We are interested in models where both the $F$ and 
$S$ fields may participate in strong-coupling dynamics, 
and hence their 
scaling properties may differ from their classical dimensions 
by potentially large anomalous dimensions.  We refer generically
to the chiral superfields of the CMSSM sector as $\phi$.

Direct couplings between the hidden and CMSSM fields arise 
from various local operators.  These are higher-dimensional operators 
and suppressed by some energy scale $M$; for the rest of this paper
we will assume that this mediation scale $M$ is at least of the order of 
the unification scale of the underlying theory (this is generically true 
in models of gravity, modulus or anomaly mediation). Analyses similar to that in this paper may be undertaken in mediation models with a lower scale, such as models of gauge mediation. 

Some of the direct interaction operators are quadratic in the hidden 
sector fields.  For example, operators that contribute to the scalar 
squared masses are
\begin{equation}
  {\cal O}_\phi : \qquad
  \int\!d^4\theta\, c^F_\phi \frac{F^\dagger F}{M^2} \phi^\dagger \phi,
\qquad
  \int\!d^4\theta\, c^S_\phi \frac{S^\dagger S}{M^2} \phi^\dagger \phi.
\label{eq:scalar-op}
\end{equation}
Other quadratic operators are
\begin{equation}
  {\cal O}_{B\mu} : \qquad
  \int\!d^4\theta\, c^F_{B\mu} \frac{F^\dagger F}{M^2} H_u H_d 
  + {\rm h.c.},
\qquad
  \int\!d^4\theta\, c^S_{B\mu} \frac{S^\dagger S}{M^2} H_u H_d 
  + {\rm h.c.},
\label{eq:Bmu-op}
\end{equation}
that contribute to the $B\mu$ parameter, the bilinear holomorphic 
supersymmetry-breaking parameter with dimension mass squared,
that appears in the Higgs sector.  Since we have scaled out powers of the mediation scale $M$, the 
coefficients $c^i$ in (\ref{eq:scalar-op}, \ref{eq:Bmu-op}) are dimensionless.
%%%%%%%%%%%%%%%%%%%%%%%%%%%%%%%%%%%

There are also operators linear in 
the hidden-sector singlet fields.  One example, for gauginos, is the mass operator
\begin{equation}
  {\cal O}_\lambda: \qquad
  \int\!d^2\theta\, c^S_\lambda \frac{S}{M} 
    {\cal W}^{a\alpha} {\cal W}^a_\alpha + {\rm h.c.},
\label{eq:gaugino-op}
\end{equation}
where the ${\cal W}^a_\alpha$ ($a=1,2,3$) are the field-strength superfields 
for the Standard Model gauge group. The operators
\begin{equation}
  {\cal O}_A: \qquad
  \int\!d^4\theta\, c^S_A \frac{S}{M} \phi^\dagger \phi + {\rm h.c.}.
\label{eq:A-op}
\end{equation}
contribute to the $A$ and $B$ parameters, the parameters associated 
with holomorphic supersymmetry-breaking scalar trilinear and bilinear 
interactions, as well as the scalar masses $|A|^2$.  And, the 
operator
\begin{equation}
  {\cal O}_\mu: \qquad
  \int\!d^4\theta\, c^S_\mu \frac{S^\dagger}{M} H_u H_d + {\rm h.c.},
\label{eq:mu-op}
\end{equation}
contributes to the $\mu$ parameter, the supersymmetric coupling
between the two observable-sector Higgs supermultiplets.
%
%%%%%%%%%%%%%%%%%%%%%%%%%%%%%%%%%%%

In the above expressions we have used the formalism of global supersymmetry; since we wish to consider high-scale mediation mechanisms such as gravity/modulus and anomaly 
mediation, we require a formulation with local 
supersymmetry.  
To convert the previous expressions to be consistent with local supersymmetry, the terms integrated over a half of the superspace 
above must then include the conformal compensator field $C$ as 
$\int\! d^2\theta\, C^3$, while the terms over the full superspace 
must include it a factor $\int\! d^4\theta\, C^\dagger C$.
%\cite{GGRS}
The latter should not be considered as part of the K\"ahler potential 
$K$, but rather as a factor in the the superspace density $f = -3 M_{\rm Pl}^2\, e^{-K/3M_{\rm Pl}^2}$, before the Weyl scaling that removes the field 
dependence in the Planck scale; $M_{\rm Pl}$ is the reduced 
Planck scale.  After Weyl scaling, each chiral superfield should be further rescaled by $1/C$ to obtain the canonical kinetic terms, 
leaving a nontrivial $C$ dependence in the various mass parameters. 
In vacua with supersymmetry breaking and no cosmological constant, 
$C = 1 + \theta^2 m_{3/2}$, where $m_{3/2}$ is the gravitino mass. 
Thus there is 
an implicit compensator dependence in all of the mass parameters, and 
sequestering (suppression) ffects occur in $f$, not in $K$.

The wavefunction renormalization factor for the operators linear in the 
hidden-sector singlet field $S$ is
\begin{equation}
  {\cal L} = \int\!d^4\theta\, Z_{S}(\mu_R)\, S^\dagger S,
\label{eq:kin-S}
\end{equation}
There are no 1PI diagrams that renormalize operators linear in $S$, 
and hence ${\cal O}_\lambda$ in Eq.~(\ref{eq:gaugino-op}), ${\cal O}_A$ 
in Eq.~(\ref{eq:A-op}), and ${\cal O}_\mu$ in Eq.~(\ref{eq:mu-op}) 
receive only the wavefunction renormalization $Z_S^{-1/2}(\mu_R)$. 
%%%%%%%%%%%%%%%%%%%%%%%%%%%%%

In general, after external-line wave-function renormalization of the 
mediation operators
linear in the hidden-sector singlet field, at an energy scale 
$\mu_R$ below the scale of hidden-sector dynamics they take the form:
\begin{equation}
  \int\!d^2\theta\, Z_S^{-1/2}(\mu_R)\, c^S_\lambda 
    \frac{S}{M} {\cal W}^{a\alpha} {\cal W}^a_\alpha + {\rm h.c.},
\label{eq:gaugino-op-Z}
\end{equation}
for the gaugino masses,
\begin{equation}
  \int\!d^4\theta\, Z_S^{-1/2}(\mu_R)\, c^S_A 
    \frac{S}{M} \phi^\dagger \phi + {\rm h.c.},
\label{eq:A-op-Z}
\end{equation}
for the $A$, $B$ parameters, and the $|A|^2$ part of the scalar 
squared masses, and
\begin{equation}
  \int\!d^4\theta\, Z_S^{-1/2}(\mu_R)\, c^S_\mu 
    \frac{S^\dagger}{M} H_u H_d + {\rm h.c.},
\label{eq:mu-op-Z}
\end{equation}
for the $\mu$ parameter.
%%%%%%%%%%%%%%%%%%%%%%%%%%%%%%%%%

Provided that there is a linear term in the superpotential,  i.e. if there is an operator
\begin{equation}
  \int\!d^2\theta\, f^2 S + {\rm h.c.},
\label{eq:S-linear}
\end{equation}
where $f$ has mass dimension one, the $S$ field acquires an $F$-component VEV.  
In the basis where the $S$ field 
is canonically normalized, this linear term is suppressed in 
the infrared as
\begin{equation}
  \int\!d^2\theta\, Z_S^{-1/2}(\mu_R)\, f^2 S + {\rm h.c.}.
\label{eq:S-linear-can}
\end{equation}
The $F$-component VEV for the canonically-normalized $S$ field is
\begin{equation}
  F_S = -Z_S^{-1/2}(\mu_R)\, f^{* 2}
\label{eq:F_S-can}
\end{equation}
and the vacuum energy is $V_0 = |Z_S^{-1/2}(\mu_R)\, f^2|^2$. Hence 
the gravitino mass is
\begin{equation}
  m_{3/2} \approx Z_S^{-1/2}(\mu_R) 
    \frac{|f|^2}{M_{\rm Pl}}.
\label{eq:m32}
\end{equation}
The wave function suppression $Z_S^{-1/2}(\mu_R)$ of Eq.~(\ref{eq:m32}), 
however, also suppresses 
all the $\mu$ and supersymmetry breaking parameters equally.  For example, we find that the 
gaugino masses are given by
\begin{equation}
  |M_a| \approx Z_S^{-1/2}(\mu_R)\, \frac{|c^S_\lambda F_S|}{M} 
  = Z_S^{-1}(\mu_R)\, 
    \frac{|c^S_\lambda f^2|}{M},
\label{eq:Ma-Z}
\end{equation}
where one factor of $Z_S^{-1/2}(\mu_R)$  in this expression comes 
from that of Eq.~(\ref{eq:m32}), and the other $Z_S^{-1/2}(\mu_R)$ 
from the suppression of the coefficient of Eq.~(\ref{eq:gaugino-op-Z}). 
This latter $Z_S^{-1/2}(\mu_R)$ provides a relative suppression 
of the gaugino masses relative to the gravitino mass: $M_a/m_{3/2} 
\sim Z_S^{-1/2}(\mu_R)$~\cite{mnp:07}.  The same is also true for the 
$\mu$ and $A$ parameters, which are also linear in $S$.  
%%%%%%%%%%%%%%%%%%%%%%%%%%%%%%%%%%%%%%%%%%%%%%%%%%%%%%%%

In fact the extra relative factor of $Z_S^{-1/2}(\mu_R)$ will appear with the corresponding 
superfield $S$ in each of the mediation operators communicating soft supersymmetry breaking 
to the observable sector.
As such, it can be affected by a constant rescaling of all the mediation operators with a factor  
$Z_S^{-1/2}(\mu_R)$ for each $S$ superfield. Since all soft masses arise from $F_{S}$,
the $S$-field F-term, this represents a homogenous rescaling of all the terms in the RGE equations 
for the soft terms. If we rescale the input soft terms at the mediation scale $M$ then, after the 
rescaling, $O_{A}$, and $O_{\mu}$, as well as $O_{\lambda}$, will all be unaffected by hidden-sector
dynamics, as the net effect has been put into this rescaling. After this rescaling, we now see that the gravitino mass has an enhancement with respect to the the gaugino mass and the  $\mu$ and $A$ parameters by a factor $Z_S^{1/2}(\mu_R)$. 
Apart from this homogenous rescaling of the input soft terms at the scale $M$, the quadratic 
operators $O_{\phi}$ and $O_{B\mu}$ also no longer receive corrections from the external line 
wave-function renormalization. However, being quadratic, the operators $O_{\phi}$ and 
$O_{B\mu}$ do receive extra renormalization due to the 1PI dynamical effects. These latter 
effects cannot be absorbed into a rescaling of the soft terms at $M$, and have observable physical consequences in the RGE flow of the soft supersymmetry-breaking parameters~\cite{crs:07}.
It is then physically important to determine the relative speed 
of suppression (sequestering) between the operators quadratic and 
linear in $S$.  Let us assume that there is no mixing between those operators 
quadratic in $S$ and those quadratic in $F$, and that only $S$ has 
a supersymmetry-breaking VEV. If there were no extra contribution 
$\alpha_S$ to the anomalous dimension of the operators from 1PI 
contributions to the hidden-sector renormalization of the operator, all 
the $B\mu$ and soft parameters would receive similar 
suppressions as $M_a \sim \mu \sim A \propto Z_S^{-1/2} F_S$ and 
$m_I^2 \sim B\mu \propto Z_S^{-1} F_S^2$, while $m_{3/2} \propto F_S$. 
Here, $m_I^2$ represent the supersymmetry breaking scalar squared 
masses.  

In general, however, the situation is more intricate~\cite{mnp:07}. 
The operators of the form ${\cal O}_\phi$ in Eq.~(\ref{eq:scalar-op}) 
and ${\cal O}_{B\mu}$ in Eq.~(\ref{eq:Bmu-op}) in general mix 
with each other, and the anomalous dimensions of the mediation operators 
bilinear in $F$, or $S$, receive non-zero corrections $\alpha_{F,S}$ from 
1PI hidden-sector renormalization.  In this case, the 
suppression of the operators quadratic in $S$ is controlled by the 
smallest eigenvalue of the $2\gamma_i \delta_{ij} + \alpha_{ij}$ 
matrix, which we define as $2\gamma_S+\hat{\gamma}_S$.  Here, $i,j$ 
runs over $F$ and $S$, and $\gamma_F$, and $\gamma_S$ are the anomalous 
dimensions of the $F$ and $S$ fields 
Note that, as we have defined it in the preceding equation, 
$\hat{\gamma}_S$ 
represents the additional renormalization-group scaling that the 
least-suppressed quadratic operators receive, relative to the operators linear in $S$,
due to their 1PI contributions, as well as their mixing with 
operators bilinear in $F$.
An additional complication is that the operators quadratic in $S$ also 
mix in a calculable way with the operators linear in $S$. In general, 
for the non-Higgs fields it is the combination $c^S_\phi 
- |c^S_A|^2$ that is suppressed by the exponent $2\gamma_S 
+ \hat{\gamma}_S$ (after potentially mixing with other quadratic 
operators), and it is this combination of operators that 
contributes to the scalar masses-squared.  Meanwhile, for the Higgs 
fields it is the combination $c^S_\phi - |c^S_A|^2 - |c^S_\mu|^2$ 
that is suppressed by the same exponent, and it is this combination of  
operators that contributes to $m_{H_{u,d}}^2 + \mu^2$.  The 
combination of operators that contributes to the $B\mu$ parameter, 
$c^S_{B\mu} - c^S_\mu (c^S_{A,H_u} + c^S_{A,H_d})$, is renormalized 
in the same way.

%%%%%%%%%%%%%%%%%%%%%%%%%%%%%%%%%%%%%%%%%%%%%%%%%%%%%%%%%%
It is precisely the extra scaling of these eigenfunction
combinations of the quadratic operators, parametrized by the 
$\hat{\gamma}_S$, whose effects we describe by general 
parametrizations in the next section, and whose effects, when combined 
with the observable-sector renormalization of the same operators, we 
show how to reconstruct from observable-sector masses and mixings.
%%%%%%%%%%%%%%%%%%%%%%%%%%%%%%%%%%%%%%%%%%%%%%%%%%%%%%%%%%

In the case that one assumes a sufficiently large interval of RGE
running to create a hierarchy between the magnitudes of the effects on the 
quadratic and the linear operators, one can obtain the following qualitatively different outcomes:
\begin{equation}
\begin{array}{lll}
  \mbox{} & M_a^2 \sim \mu^2 \sim A^2 
    \gg m_{Q_i,U_i,D_i,L_i,E_i}^2 \sim B\mu \sim m_{H_{u,d}}^2 + \mu^2 
    & (\hat{\gamma}_S > 0),
\\
  \mbox{} & M_a^2 \sim \mu^2 \sim A^2 
    \ll m_{Q_i,U_i,D_i,L_i,E_i}^2 \sim B\mu \sim m_{H_{u,d}}^2 
    & (\hat{\gamma}_S < 0),
\end{array}
\end{equation}
%modified
%
depending on the sign of the exponent $\hat{\gamma}_S$.  (In the 
absence of the operator mixing, $\hat{\gamma}_S = \alpha_S$.)  
In addition, since the gravitino mass is generally enhanced by wave-function renormalization effects relative to all the soft parameters, it is also possible that anomaly mediation contributions may be a significant contribution to the low-energy soft supersymmetry-breaking parameters, giving rise to a mirage pattern of soft scalar mass terms.
%(Case~3).  
Although it is not possible to work out the signs or 
magnitudes of the exponents $\hat{\gamma}_S$ 
for a given strongly-coupled 
theory with the currently 
available technology, theories of this class may be parametrized by their 
$\hat{\gamma}_S$ values. In the next section we consider simple general 
parametrizations of the strongly-coupled hidden-sector dynamics
in terms of these exponents 
$\hat{\gamma}_S$. The primary goal of this paper is to demonstrate that, in 
supersymmetric unified theories with strongly-coupled hidden sectors, 
the parameters and type of the hidden sector can be inferred from low-energy observables, 
at least for hidden sectors with simple two-parameter 
parametrizations.
%%%%%%%%%%%%%%%%%%%%%%%%%%%%%%%%%%%%%%%%%%%%%%%%%%%%%%%%%%

Finally, we note that, whilst the 
quadratic operator 1PI renormalizations affect directly only the soft 
supersymmetry-breaking scalar mass-squared $|M_{\phi}|^{2}$ and the $B\mu$ parameters, 
under the RGE evolution due to 
observable-sector interactions these terms feed back into the other soft 
supersymmetry-breaking mass terms such as gaugino masses, at higher 
order.

%%%%%%%%%%%%%%%%%%%%%%%%%%%%%%%%%%%%%%%%%%%%%%%%%%%%%%%
%%%%%%%%%%%%%%%%%%%%%%%%%%%%%%%%%%%%%%%%%%%%%%%%%%%%%%%
%%%%%%%%%%%%%%%%%%%%%%%%%%%%%%%%%%%%%%%%%%%%%%%%%%%%%%%
%%%%%%%%%%%%%%%%%%%%%%%%%%%%%%%%%%%%%%%%%%%%%%%%%%%%%%%

\section{Hidden-Sector Behaviour: General Parametrizations}

As we discussed in the last section, after diagonalizing the anomalous-dimension  mixing matrix for the quadratic hidden-sector scalar mass-squared terms, we find that the 
hidden-sector interactions responsible for generating the scalar and gaugino masses of the CMSSM take the generic forms, 
\be
\label{non_ren}
\int d^4\theta\hspace{1mm} k_i \frac{X^\dagger X}{M^2} \phi_{i}^\dagger \phi_{i} + \int d^2\theta \hspace{1mm} c^S_\lambda \frac{S}{M} 
    {\cal W}^{a\alpha} {\cal W}^a_\alpha + {\rm h.c.},
\ee
%%%%%%%%%%%%%%%%%%%%%%%%%%%%%%%%%%%%%%%%%%%%%%%%%%%%%%%%%%%%%%%%%%%%%%%
%%%%%%%%%%%%%%%%%%%%%%%%%%%%%%%%%%%%%%%%%%%%%%%%%%%%%%%%%%%%%%%%%%%%%%%
%%%%%%%%%%%%%%%%%%%%%%%%%%%%%%%%%%%%%%%%%%%%%%%%%%%%%%%%%%%%%%%%%%%%%%%
 \begin{figure}[ht!]
    \newlength{\picwidthaa}
    \setlength{\picwidthaa}{2.7in}
    \begin{center}
 	   \subfigure[][]{\resizebox{\picwidthaa}{!}{\includegraphics{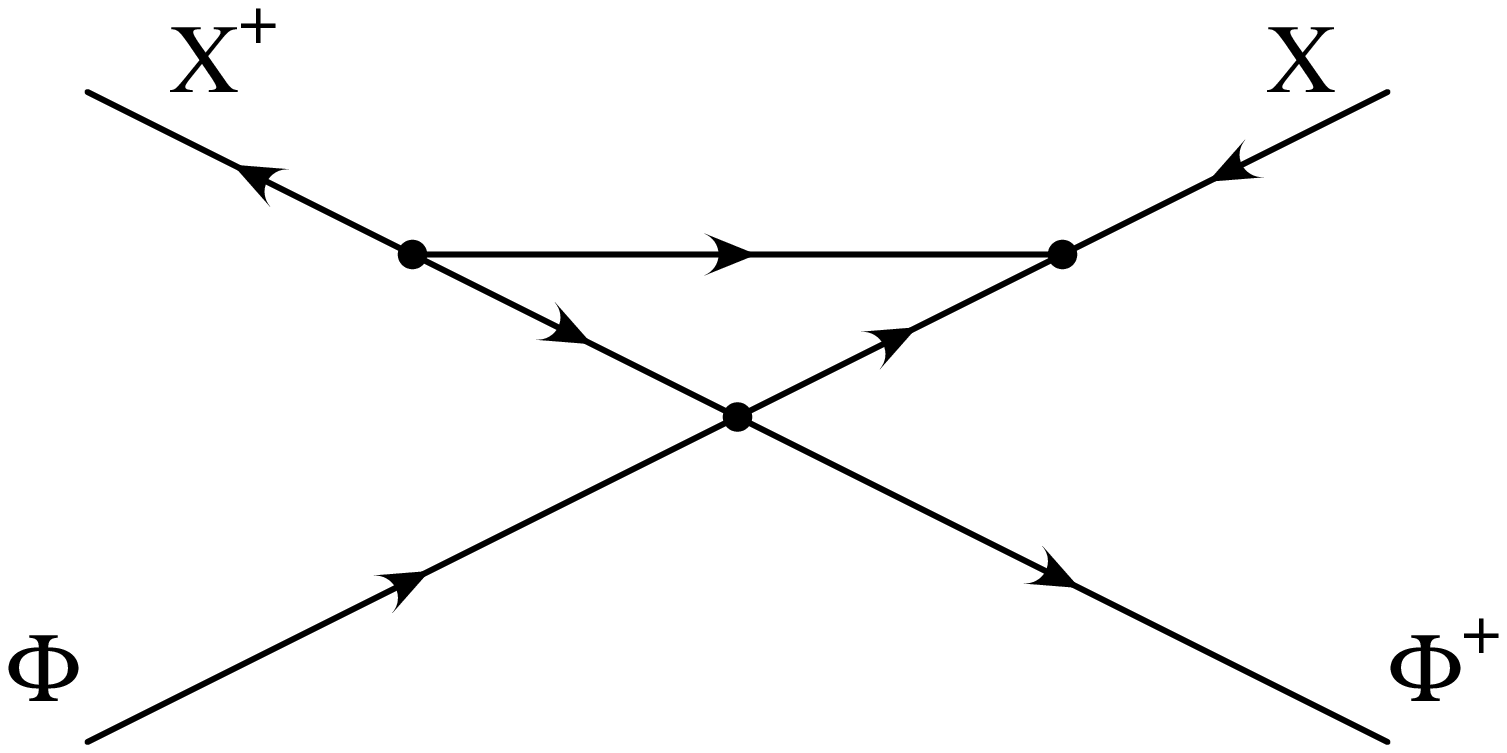}}}
 	   \subfigure[][]{\resizebox{\picwidthaa}{!}{\includegraphics{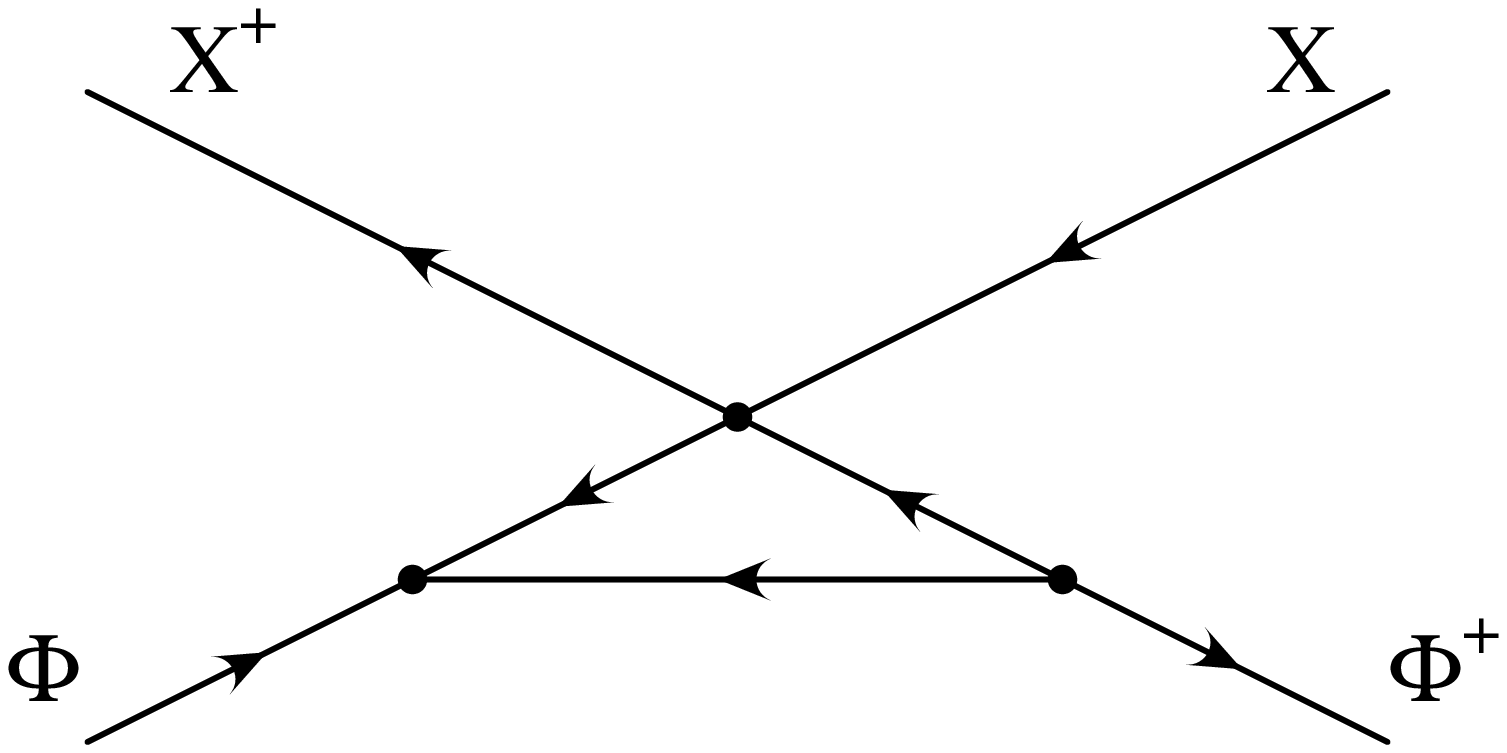}}}
        	   \subfigure[][]{\resizebox{\picwidthaa}{!}{\includegraphics{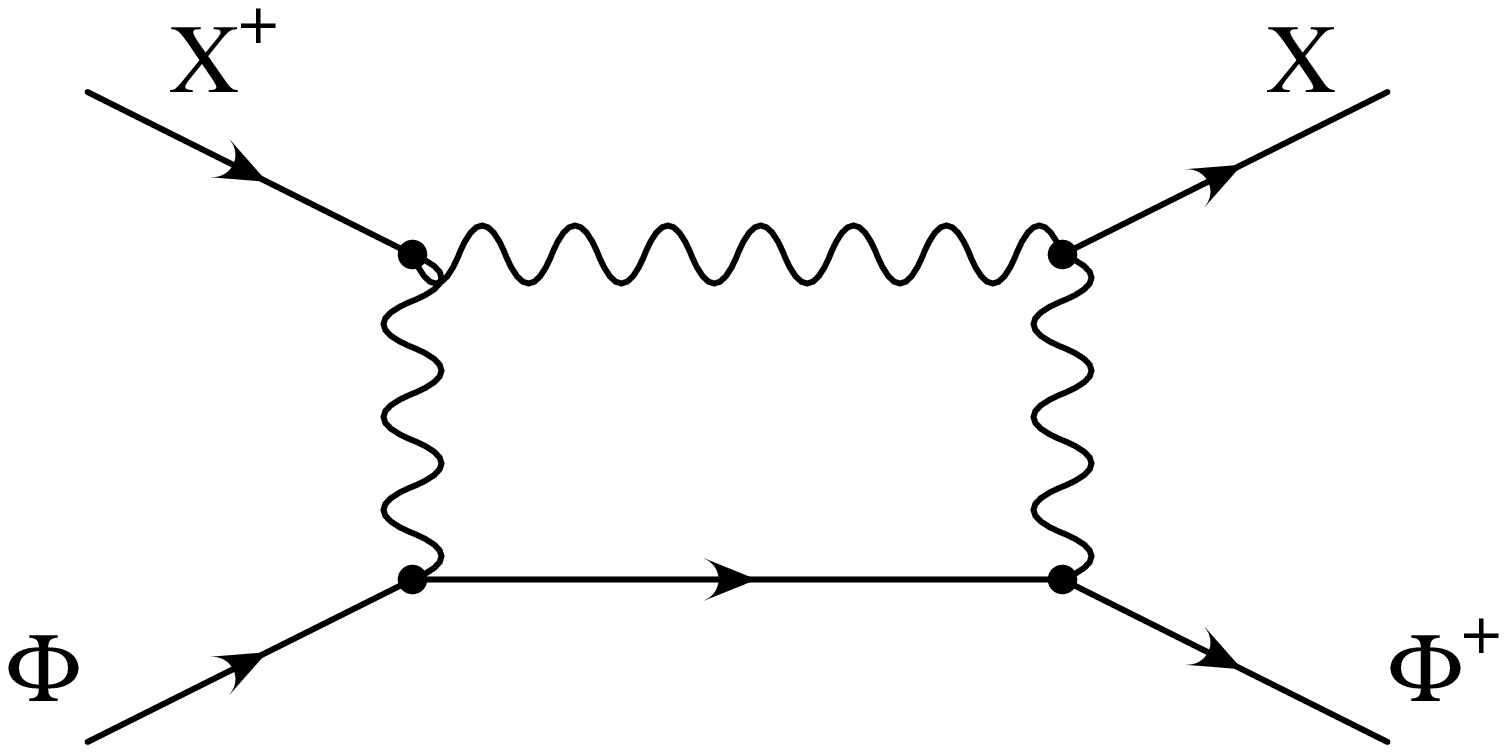}}}
    \end{center}
    \caption{\it One-loop supergraphs contributing to the soft supersymmetty-breaking scalar mass renormalization.}
    \label{super_g}
 \end{figure}
%%%%%%%%%%%%%%%%%%%%%%%%%%%%%%%%%%%%%%%%%%%%%%%%%%%%%%%%%%%%%%%%%%%%%%%
%%%%%%%%%%%%%%%%%%%%%%%%%%%%%%%%%%%%%%%%%%%%%%%%%%%%%%%%%%%%%%%%%%%%%%%
where $X$ denotes the hidden-sector field corresponding to the dominant 
eigencombination of the $F$ and $S$, generically the one with the smallest 
$\hat{\gamma}_S$~\footnote{In exceptional cases, if the 1PI hidden-sector renormalization is weak 
and there is a larger initial value for a 
linear eigencombination corresponding to a larger value of 
$\hat{\gamma}_S$, the large initial value may overwhelm the scaling suppression, implying 
that the more suppressed eigencombination survives as the dominant eigencombination, in which 
case it would be that combination we label as $X$.}, $k_{i}$ corresponds to the combination of 
$c_{\phi}^{F}$ and $c_{\phi}^{S}$ corresponding to the linear combination of $F$ and $S$ 
that composes the dominant eigencombination $X$ (technically it is the largest  $k_{i}X^\dagger X$ 
that determines which is the dominant eigencombination for the purposes of keeping the 
dominant operator whose scaling we wish to consider),
$M$ represents the messenger scale, and $\phi_{i}$ and 
${\cal W}^a_\alpha$ denote the visible-sector scalar and gaugino fields. 
As we see in Fig.~\ref{super_g}, both the hidden- and visible-sector interactions renormalize 
the coefficients $k_i$ and, since the $k_i$ coefficients are responsible for setting the 
observable-sector soft supersymmetry-breaking scalar masses-squared in the CMSSM, the 
hidden sector can have an important impact on the final spectrum of scalars predicted
at low energies. As a simple illustrative and analytic example, the 
one-loop contribution to the renormalization of $k_i$ involving the hidden and gauge sectors 
yields~\cite{crs:07} (we consider observable-sector superpotential interactions later),
\be
\frac{d}{dt}k= \gamma(t) k_i - \frac{1}{16\pi^2} \sum_n 8 C_2^n(R_i) g^6_n(t) G ,
\label{RGeq}
\ee
where $\gamma(t)$ ($\equiv \hat{\gamma}_S(t)$) denotes the anomalous dimension arising from the
hidden-sector interactions in a hidden-sector renormalization scheme where we have removed the external-line wave-function renormalization of the $S$ field by a rescaling of the soft 
supersymmetry-breaking mass terms, as discussed in the previous section, and
the second term represents the leading visible-sector contribution containing the 
representation-dependent factors $C_2(R_i)$ and gauge couplings, $g_n$. The solution to 
eq.(\ref{RGeq}) is
\be
k_i(t) = \exp\left(-\int_t^0 dt^\prime \gamma(t^\prime)\right) k_i(0) + \frac{1}{16\pi^2}\sum 8 C_2^n(R_i) \int_t^0 ds g^6_n(s) \exp\left(-\int_t^s dt^\prime \gamma(t^\prime)\right) G.
\label{RGsol}
\ee
At this point, if we wish to study a particular model of the hidden sector, we may simply input it into the solution eq.(\ref{RGsol}) along with the running of the observable-sector gauge couplings. 

However, it is useful to parametrize the general types of behaviour that may arise in 
hidden-sector effects on the RGE flow of the scalar masses. Since  we wish to infer the nature of 
the hidden sector from the indirect effect it has on the flow of observable sector masses, through 
hidden-sector contributions to the operator RGE flow, in the presence of large observable-sector 
RGE contributions (gauge couplings, $t$ Yukawa, possibly $b$ and $\tau$ Yukawas), we would 
expect that, in order for the hidden-sector effects to be noticeable, the hidden-sector coupling 
should become large somewhere in the range of running. Generally, one then expects a large contribution either at the lower range of running (near $M_{hid}$), or at the upper range of running 
(near $M$, the messenger scale), or an approximately constant (scale-invariant) contribution
over the range between $M_{hid}$ and $M$. Furthermore, the sign of the effect on the operator 
will, in general, not be correlated with the apparent $\beta$-function behaviour corresponding to 
growth in the ultraviolet or infrared, because the apparent running may be a product of the 
operator mixing that we discussed in the previous section. Each of these cases may be 
described approximately by a simple two-parameter parametrization first proposed in~\cite{ros:07}, 
which expresses $\gamma(t)$ as
\be
\gamma(t) = \frac{1}{b_\gamma(t-a_\gamma)}
\ee
where $t= \ln(\mu/M)$. This is similar to how we parametrize the observable-sector gauge 
interactions:
\be
g_n^2(s) = \frac{1}{b_n(s-a_n)} 
\ee
where, for example,  $a_s = \log(\Lambda_{QCD}/M)$ would yield the standard one-loop
parametrization of the QCD coupling. This two-parameter 
($b_\gamma$ and $a_\gamma$) representation of the hidden-sector renormalization effects 
permits a simple mathematical characterization of the full class of possible model behaviours 
described above (models with approximately scale-invariant effects from the hidden sector 
over the range of running have an even simpler parametrization, as we discuss below).
For example, plugging into the general solution of our simple analytic example, we find that 
\be
\exp\left(-\int_t^s dt^\prime \gamma(t^\prime)\right) = \left(\frac{t-a_\gamma}{s-a_\gamma}\right)^{1/b_\gamma},
\ee
which allows us to write
\be
k_i(t) = (t-a_\gamma)^{1/b_\gamma}\left[(-a_\gamma)^{-1/b_\gamma} k_i(0) + \frac{1}{16\pi^2} \sum_n 8 C_2^n(R_i)G\int_t^0 ds\left(\frac{1}{b_n(s-a_n)}\right)^3 (s-a_\gamma)^{-1/b_\gamma}\right].
\ee
We are now in a position to study the qualitative behaviours of the different cases parametrized above. 
\\
{\bf Case 0:} $\gamma = 0$.\\
This trivial case corresponds to a negligible hidden-sector effect. Only the usual visible sector renormalization of the CMSSM appears.
\\
{\bf Case 1:} $b_\gamma = -1$, $ a_\gamma =1$.\\ 
For the case of $b_\gamma = -1$, the integral can be explicitly done and the solution becomes:
\be
k_i(t) = (t-a_\gamma)^{-1}\left[-a_\gamma k_i(0) + \frac{1}{16\pi^2} \sum_n 8 C_2^n(R_i)\frac{G}{b^3_n}\left(\frac{a_n+a_\gamma}{2 a_n^2} + \frac{1}{t-a_n} - \frac{a_\gamma - a_n}{2(t-a_n)^2}\right) \right].
\ee
With $a_\gamma =1$ this case corresponds to IR-free renormalizations in the hidden sector with strong coupling effects
(perhaps nonperturbative) at mediation scale $M$. The choice 
$b_\gamma = -1$ represents an extreme situation in the sense that the evolution becomes slow (walking), which enhances the hidden-sector effect. 
%Recall that in QCD, $b_n = 2(11-2n_f/3)$. 
This gives an IR mass suppression at $t_{hid} = \ln(M_{hid}/M)$ (with the suppression occuring predominantly at scales near $M$) and can give rise to ``mirage unification'',
in which the scalar masses appear to unify at an intermediate scale. We return to this case in 
detail in Section~\ref{example_im}.
\\
{\bf Case 2:} $b_\gamma =-1$, $a_\gamma = \log(M_{hid}/M)$. \\
The analytic solution in this case is the same as in the preceding 
$b_\gamma = -1$ case. But now because $a_\gamma = \ln(M_{hid}/M)$
we have UV-free effective renormalizations in the hidden sector with effects that become nonperturbative at $eM_{hid}$.
At the scale $t_{np} = \ln(eM_{hid}/M)$ we get an IR mass enhancement (with the enhancement occuring predominantly at scales near $M_{hid}$). In this case the unification of masses 
may become completely obscured as they ''blow apart''.
\\
{\bf Case 3:} $b_\gamma =1$, $a_\gamma = 1$. \\
In the case $b_\gamma = 1$, the integral can be done explicitly, and the solution becomes:
\bea
k_i(t) && = (t-a_\gamma)\left[(-a_\gamma)^{-1} k_i(0) + \frac{1}{16\pi^2}\sum_n 8 C_2^n(R_i) \frac{G}{b_n^3}\left[\frac{1}{(a_\gamma -a_n)^3}\log\left(\frac{a_\gamma(t-a_n)}{a_n(t-a_\gamma)}\right)\right.\right. \nonumber \\
&& \left.\left. + \frac{1}{(a_\gamma-a_n)^2}\left(\frac{-1}{a_n} - \frac{1}{t-a_n}\right) + \frac{1}{2(a_\gamma - a_n)}\left(\frac{1}{a_n^2} - \frac{1}{(t-a_n)^2}\right)\right]\right].
\eea
With $a_\gamma =1$ this case corresponds to IR-free renormalization in the hidden sector with 
strong-coupling effects
(possibly nonperturbative) at the mediation scale $M$. Because of the negative sign of $\gamma(t)$ 
in the regime of running, the masses are IR-enhanced (predominantly from scales near $M$), which means that unification of masses could become obscured as they ''blow apart''.
\\
{\bf Case 4:} $b_\gamma =1$, $a_\gamma = \ln(M_{hid}/M)$. \\
The analytic solution in this case is the same as in the preceding 
$b_\gamma = 1$, case. But now because $a_\gamma = \ln(M_{hid}/M)$
we have UV-free effective renormalizations in the hidden sector with effects that become nonperturbative at $M_{hid}$. 
This gives an IR mass suppression at $t_{hid} = \log(M_{hid}/M)$ (with the suppression occuring predominantly at scales near $M_{hid}$) and can give rise to ``mirage unification'',
in which the scalar masses appear to unify at an intermediate scale.
\\

For hidden sectors with approximately scale-invariant behaviour (somewhere) in the range 
between $M_{hid}$, the scale of spontaneous supersymmetry breaking in the hidden sector, 
and $M$, the mediation scale, it is easy to find simple two-parameter characterizations of 
their behaviour. One parameter is clearly the (constant) value of $\gamma$ over the scaling interval. 
The other parameter will govern the extent of the scaling interval; this will be a mass scale $M_{c}$ 
that governs the limit of the region of approximately scale-invariant running. The scale-invariant 
region could be at the high end of the interval, starting at $M$ and running down to $M_{c}$, or at the low end of the interval, starting at 
$M_{c}$ and running down to $M_{hid}$. We leave to readers the exercise of working
out the solution of our simple analytic example for hidden sectors parametrized by these behaviours.

With simple parametrizations of the potential behaviours of the hidden-sector contributions to the 
RGE running of the scalar masses, we may now turn to the question of their effect on the 
soft supersymmetry-breaking terms that could be measured at observable energies. 
Appendix A lists sum rules between the scalar masses that hold even in the presence of
hidden-sector effects (see also~\cite{kkm:08}).
In general, one needs to integrate numerically the full set of MSSM RGE equations 
(see Appendix B for these equations at one-loop order) with the addition of the hidden-sector 
scalar mass-squared anomalous-dimension terms as in (\ref{RGeq}), in a parametrization such as 
those proposed in this Section. In the next Section we consider the degree to which the 
resulting predictions for the soft supersymmetry-breaking parameters may be used to determine 
the type and parameters of the hidden sector that induced them, rendering the hidden sector
`observable'.

%%%%%%%%%%%%%%%%%%%%%%%%%%%%%%%%%%%%%%%%%%%%%%%%%%%%%%%%%%
%%%%%%%%%%%%%%%%%%%%%%%%%%%%%%%%%%%%%%%%%%%%%%%%%%%%%%%%%%
%%%%%%%%%%%%%%%%%%%%%%%%%%%%%%%%%%%%%%%%%%%%%%%%%%%%%%%%%%
%%%%%%%%%%%%%%%%%%%%%%%%%%%%%%%%%%%%%%%%%%%%%%%%%%%%%%%%%%

\section{Hidden-Sector Reconstruction}
\label{example_im}

In this Section we consider how the effects of the hidden-sector renormalization, combined with that 
from observable-sector interactions,
can cause observable distortions in the low-energy spectrum of the soft supersymmetry-breaking 
scalar mass-squared terms. We make minimal assumptions on the spectrum of the soft 
supersymmetry-breaking scalar mass-squared terms input at the unification scale.
As a first example we assume a universal value for squark and slepton masses squared.
Subsequently we relax this assumption, but keeping a common value for scalars with the same 
gauge charges, as seems to be required by the limits on neutral supersymmetric flavour 
violation~\cite{EN,BG,il:02,cam:02}.
We show that, with these assumptions, we can in principle distinguish the general classes of 
two-parameter hidden-sector parametrizations introduced in the last Section, and
reconstruct their parameters.

We illustrate this reconstruction using a two-parameter hidden sector of the type of Case 1 from the previous Section~\footnote{Readers can easily adapt our arguments to all the parametrizations of the different behaviours described in the previous section. Indeed, in a companion paper~\cite{cem:2} we study neutrino seesaw physics with both this and several alternative parametrizations of hidden-sector behaviours.}. In order to facilitate the comparison to previous literature, we describe 
it differently. Following~\cite{din:04,crs:07} we consider a toy self-interacting hidden sector that 
contains the superpotential term
\be
\label{Inter}
W_{\mathrm{h}} = \frac{\lambda}{3!} X^3.
\ee
This simple superpotential by itself does not break supersymmetry, and hence the hidden sector of 
eq.(\ref{Inter}) must be enlarged by additional interactions responsible for generating F- or D-term 
VEVs in a realistic model. For the purposes of examining the effects of hidden-sector 
renormalization, following~\cite{din:04,crs:07} we suppose that eq.(\ref{Inter}) appears as the 
dominant self-interaction term in the hidden-sector superpotential, and that it provides the 
dominant hidden-sector contribution to the anomalous dimension of the operator 
mediating supersymmetry breaking scalar masses in the observable sector. 
The observable interactions generate the usual MSSM RGEs, while the separate hidden-sector renormalizations of the coefficients $k_i$ of eq.(\ref{non_ren}) -- resulting from the self-interactions 
in eq.(\ref{Inter}) -- add an additional contribution. The hidden-sector Yukawa interactions at 
lowest order in $\lambda$ yield
$\gamma(t) = (2\lambda^*(t)\lambda(t))/(16\pi^2)$ where, as per our convention, we have removed 
the external-line wave-function contributions to the RGE runnning or the operator, and  
$\lambda(t)$ is the running hidden-sector Yukawa coupling, which to lowest order satisfies:
\be
\frac{d\lambda}{d t} = \frac{3}{32\pi^2} \lambda^3.
\ee

In line with the aim of this paper, and to facilitate comparison to the previous literature, 
we now define our (two-parameter) parametrization of this hidden sector. We use as our two 
parameters the value of the coupling $\lambda$ at the unification scale and the value of the scale 
$M_{hid}$ which gives the infrared cutoff on the hidden-sector dynamical scale.
We consider this as variable, yielding a two-parameter toy model of the hidden sector.
Finally, we define the leading-order expression 
$\gamma(t) = (2\lambda^*(t)\lambda(t))/(16\pi^2)$ as the exact value of the anomalous 
dimension $\gamma(t)$. This is clearly true only at leading order in the theory we wrote down 
in eq.(\ref{Inter}), but it defines a convenient parametrization of the behaviour we wish to assume 
in our example, and it facilitates comparison to previous results which use the leading order 
expression. Similarly, for our numerical studies we use the leading-order RGE behaviour of the 
coupling $\lambda$ as if it were exact. 
This yields a hidden-sector contribution to the running of $k_{i}$ given by
\be
\frac{d k_i}{dt} = \frac{2\lambda^*\lambda}{16\pi^2} k_i ,
\ee
implying that the scalar mass RGEs of the MSSM become augmented between the 
unification and messenger scales, becoming:
\be
\frac{d{\bf m_S}^2}{dt} \rightarrow \frac{d{\bf m_S}^2}{dt} + \frac{2\lambda^*\lambda}{16\pi^2}{\bf m_S}^2.
\label{aug}
\ee
Again remembering our convention that we remove the external-line wave-function
renormalization effect (i.e., we work as if the gaugino masses are held fixed under 
hidden-sector renormalization). We see in eq.(\ref{aug}) that the hidden-sector contribution 
is the same for all scalar mass-squared RGEs, and therefore the effect on the running of the
scalar masses-squared depends on the representation of the scalar particle itself. This effect 
suggests a fully general strategy for uncovering the effects of the hidden sector
on the low-energy spectrum. 

We demonstrate this strategy with two examples, and then detail the general method of hidden-sector reconstruction. In our two examples, we consider the cubic superpotential for the hidden sector as 
given in 
eq.(\ref{Inter}), yielding an infrared-free hidden sector $\beta$-function for the Yukawa coupling. 
We stress that this choice of the hidden sector represents a proof-of-principle example, chosen 
to facilitate comparison with the existing literature, and should not be considered as a
realistic scenario. 

\subsection{Example 1: Universal Scalar Masses and Observable-Sector Gauge Interactions}

In this first example, we also assume universal scalar masses and 
universal gaugino masses at the unification scale $M_{X} \approx 2\times 10^{16}$ GeV, 
and an intermediate hidden-sector scale. Also, we assume that all trilinear A-terms vanish at 
$M_{Pl}$, and we assume a positive sign for $\mu$. Given these model assumptions, we now demonstrate that it is possible to reconstruct the parameters of the hidden sector using the 
observable low-energy scalar masses. 

\begin{figure}[ht!]
    \newlength{\picwidtha}
    \setlength{\picwidtha}{6.2in}
    \begin{center}
        \resizebox{\picwidtha}{!}{\includegraphics{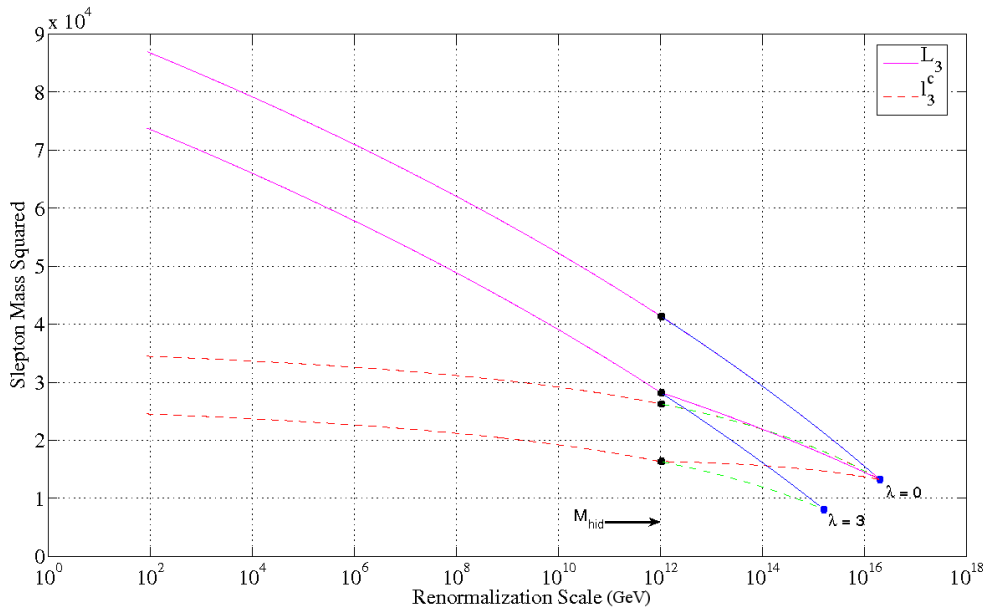}}
    \end{center}
    \caption{\it The RGE flows of the largest mass-squared eigenvalues for the third-generation
    left $(L_3)$ and 
    right-handed $(l^c_3)$slepton for $\lambda =0$ and $\lambda =3$.}
    \label{Mirage_1}
 \end{figure}

Since the effects of the hidden sector on the scalar masses-squared depend on the 
representations of the scalar particles, we first plot the running for two different species with the 
hidden sector turned on, and then extrapolate the low-energy result backward with the 
hidden sector turned off, so as to examine fully its effects. 
We see some typical results in Fig.~\ref{Mirage_1},
where we have assumed a universal scalar mass of $115$~GeV and a universal gaugino mass of 
$375$~GeV at $M_{X}$, and we have placed the hidden-sector scale $M_{hid}$ 
(where we integrate the hidden sector out of the system) at $10^{12}$ GeV. 
In Fig.~\ref{Mirage_1} we run downward the RGEs of the masses-squared of the third-generation 
left-handed slepton, $L_3$, and of the third-generation right-handed slepton, 
$l^c_3$, using universal boundary conditions at $M_X = 2 \times 10^{16}$~GeV, 
to yield low-energy predictions for 
the scalar masses-squared. The uppermost solid and dashed lines in Fig.~\ref{Mirage_1} show
the RGE evolution if the hidden-sector coupling $\lambda$ vanishes. A naive extrapolation up
from low energies using just the MSSM RGEs yields the correct result that the masses unify at $M_X$.

However, since the left- and right-handed sleptons sit in different representations of 
SU(2)$_L$, the RGE flows of the two scalar masses proceed differently between $M_X$ and 
$M_{hid}$ for non-zero $\lambda$, as shown by the middle solid and dashed lines in 
Fig.~\ref{Mirage_1} that are present between $M_X$ and $M_{hid}$. 
Using the low-energy predictions obtained from the theory 
with the hidden-sector effects incorporated to define the initial conditions at low energy, if we
naively run the MSSM RGEs upward, neglecting the effect of the hidden sector (as shown by the lowest
solid and dashed lines in Fig.~\ref{Mirage_1}), we see that the intersection 
of the $L_3$ and $l^c_3$ masses appears to take place at a scale different from $M_{X}$ if
$\lambda$ is non-zero. That is, the hidden-sector effect predicts that the RGE flow of the usual 
MSSM applied naively to the observed low-energy scalar masses-squared of different representations will appear to yield ``mirage unification" at a scale distinct from the gaugino unification point at 
$M_{X}$ (which the hidden-sector effect leaves unaltered). The appearance of such a mirage 
scalar-unification 
scale would be the first indication of the presence of a hidden-sector effect~\footnote{For a
phenomenological analysis of `mirage' scenarios, see~\cite{Baer}.}. 

\begin{figure}[ht!]
    \newlength{\picwidthb}
    \setlength{\picwidthb}{6.2in}
    \begin{center}
        \resizebox{\picwidthb}{!}{\includegraphics{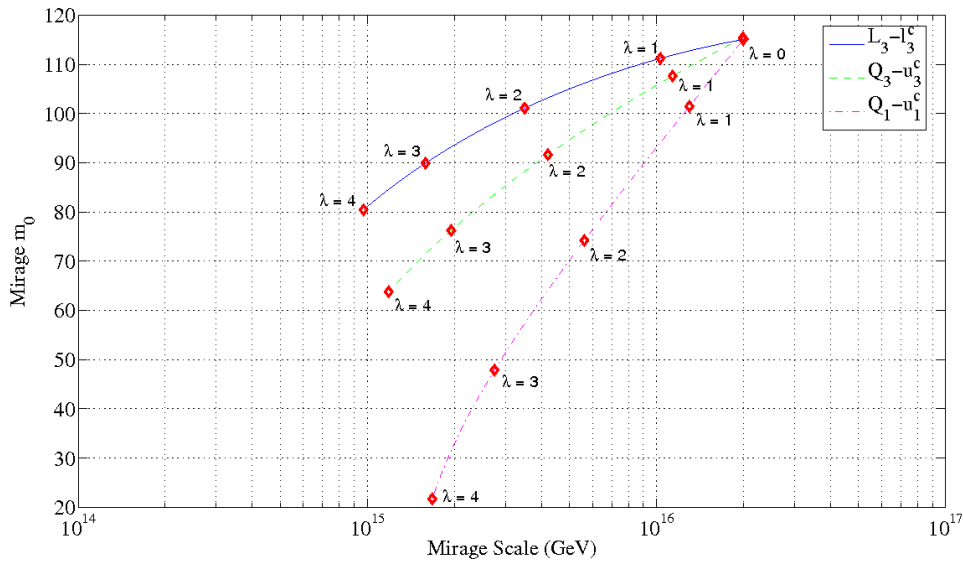}}
    \end{center}
    \caption{\it The dependences of the mirage scale and the mirage scalar mass on $\lambda$ for
    $M_{\mathrm{hid}} = 10^{12}$ GeV, using the same MSSM inputs as in Fig.~\protect\ref{Mirage_1}.}
    \label{Mirage_2}
 \end{figure}

In Fig.~\ref{Mirage_2} 
we plot the mirage scale as a function of predicted scalar mass at the low scale. Here we have used backward extrapolation to determine the intersection points for the pairs ($L_3$,$l_3$), 
($Q_3$,$u^c_3$), and ($Q_1$,$u^c_1$) for $\lambda=0$--$4$. In all cases we have used the same 
MSSM inputs as in Fig.~\ref{Mirage_1}.
We emphasis again that the mirage scale is a fake, in that it simply indicates the presence of 
additional structure relative to the pure weakly-coupled CMSSM scenario.

While Figs.~\ref{Mirage_1} and \ref{Mirage_2} demonstrate how a mirage scale develops, by 
examining the RGE flow using the low-energy output of a particular high-energy 
implementation, empirically we will only have the low-energy spectrum -- we will not know the scale 
at which to integrate out the hidden sector, nor will we know its Yukawa coupling $\lambda$. 
Thus, we require a complete bottom-up approach and an understanding of the level of 
parameter degeneracy in this system. 

Under the assumption of universal scalar masses and a 
hidden sector given by eq.(\ref{Inter}), a fixed universal scalar mass at $m_0$ can lead to the same 
prediction for the low-energy scalar mass for different values of $M_{hid}$ and $\lambda$. 
Furthermore, different values of $m_0$ can lead to the same low-energy prediction by 
compensating different inputs with $M_{hid}$ and $\lambda$~\footnote{Added complications 
appear if we do not assume total scalar universality, and we address this issue in our second example.}. 
%We can see these parameter degeneracies more clearly in figure \ref{par_den}.
 \begin{figure}[ht!]
    \newlength{\picwidthc}
    \setlength{\picwidthc}{6.2in}
    \begin{center}
        \resizebox{\picwidthc}{!}{\includegraphics{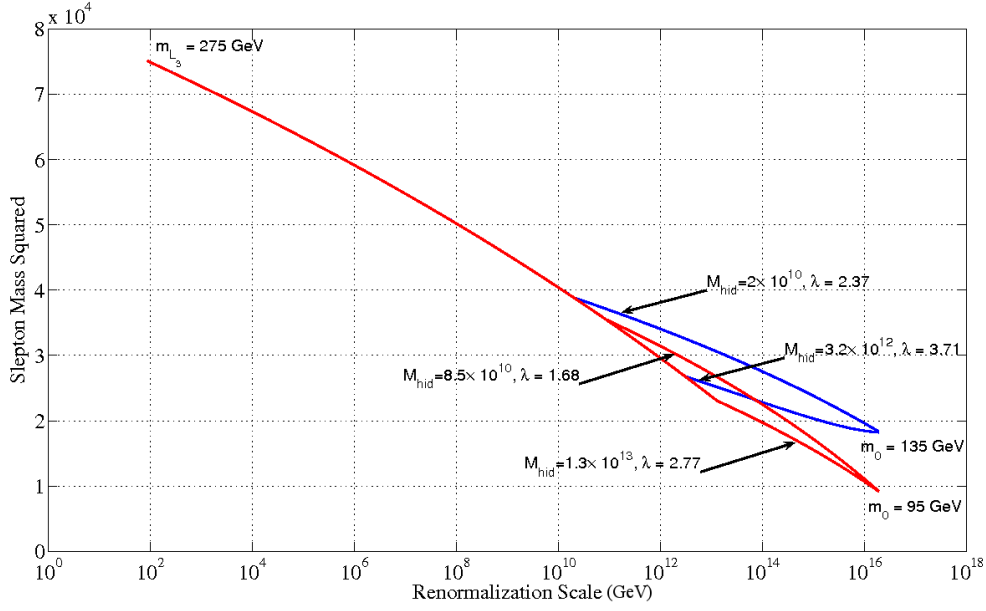}}
    \end{center}
    \caption{\it Different choices of the hidden-sector parameters and the universal input scalar mass 
    $m_0$ can lead to the same low-energy prediction, $m_{L_3} = 275$~GeV in this case.}
    \label{par_den}
 \end{figure}
Fig.~\ref{par_den} demonstrates four different parameter choices that lead to the same predicted 
value for the largest left-handed slepton eigenvalue, $m_{L_3} = 275$~GeV at the weak scale. 
This figure indicates that contours of constant $m_0$ develop in the $M_{hid}$ -- $\lambda$ plane,
that lead to the same low-energy scalar mass prediction.
 \begin{figure}[ht!]
    \newlength{\picwidthd}
    \setlength{\picwidthd}{6.2in}
    \begin{center}
        \resizebox{\picwidthd}{!}{\includegraphics{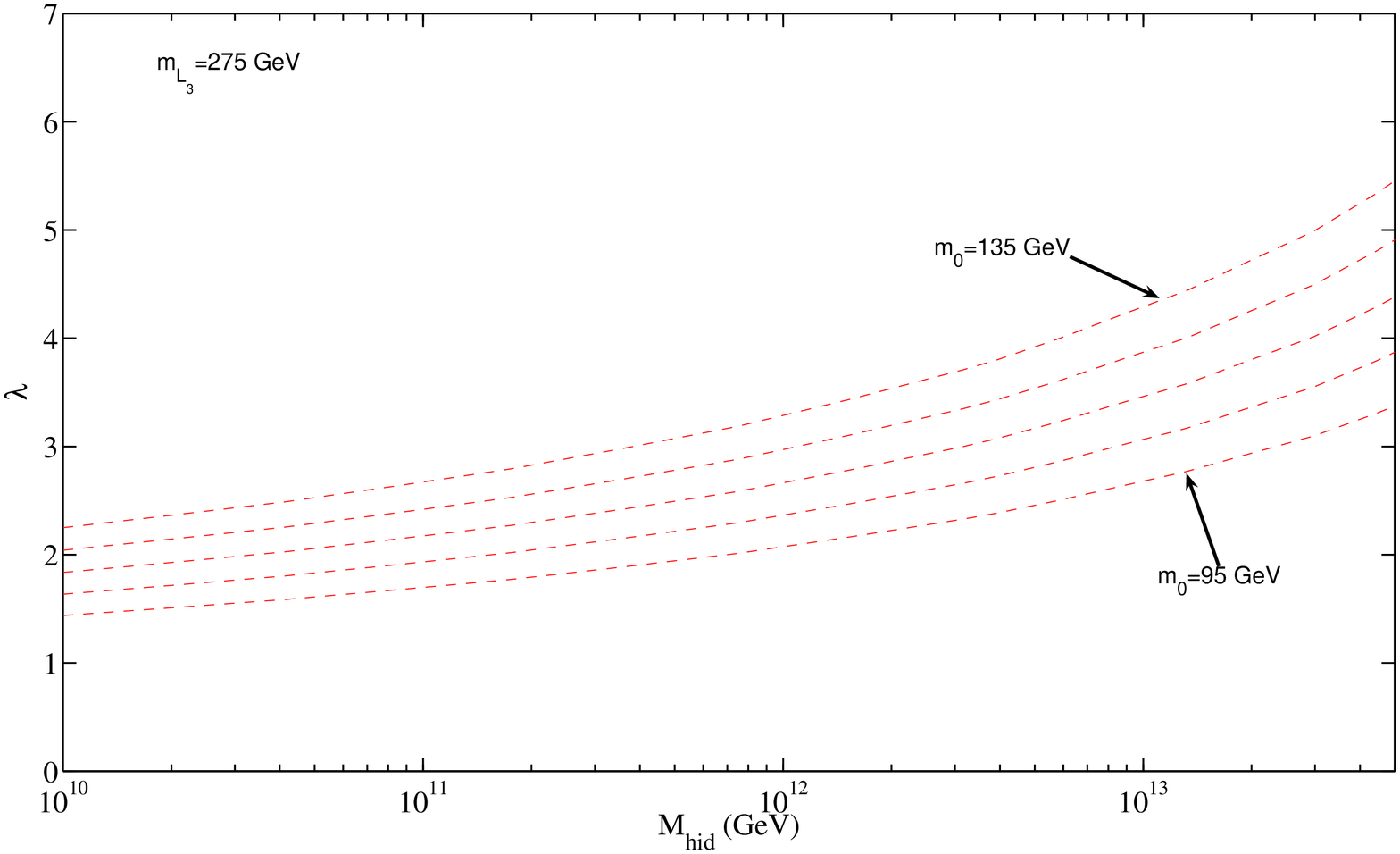}}
    \end{center}
    \caption{\it For a fixed low-energy prediction of a slepton mass ($m_{L_3} = 275$~GeV in this case), 
    contours at different $m_0$ exist in the $M_{hid}$--$\lambda$ plane that predict the 
    same low-energy mass value.}
    \label{contour_indy}
 \end{figure}
%We can see this effect in figure \ref{contour_indy}. 

In Fig.~\ref{contour_indy} we show the effects of allowing $M_{hid}$ and $\lambda$ to vary 
while ensuring the same low-energy prediction of Fig.~\ref{par_den}, namely 
$m_{L_3}=275$~GeV at the weak scale. Allowing $m_0$ to range from $95$~GeV to $135$~GeV in 
$10$~GeV increments, we produce five contours in the $M_{hid}$--$\lambda$ plane as indicated 
by the dashed line in Fig.~\ref{contour_indy}. We see that information on $m_{L_3}$
alone is insufficient to
determine the value of $m_0$ or the hidden-sector parameters from low-energy data. 

\begin{figure}[ht!]
    \newlength{\picwidthe}
    \setlength{\picwidthe}{6.2in}
    \begin{center}
        \resizebox{\picwidthe}{!}{\includegraphics{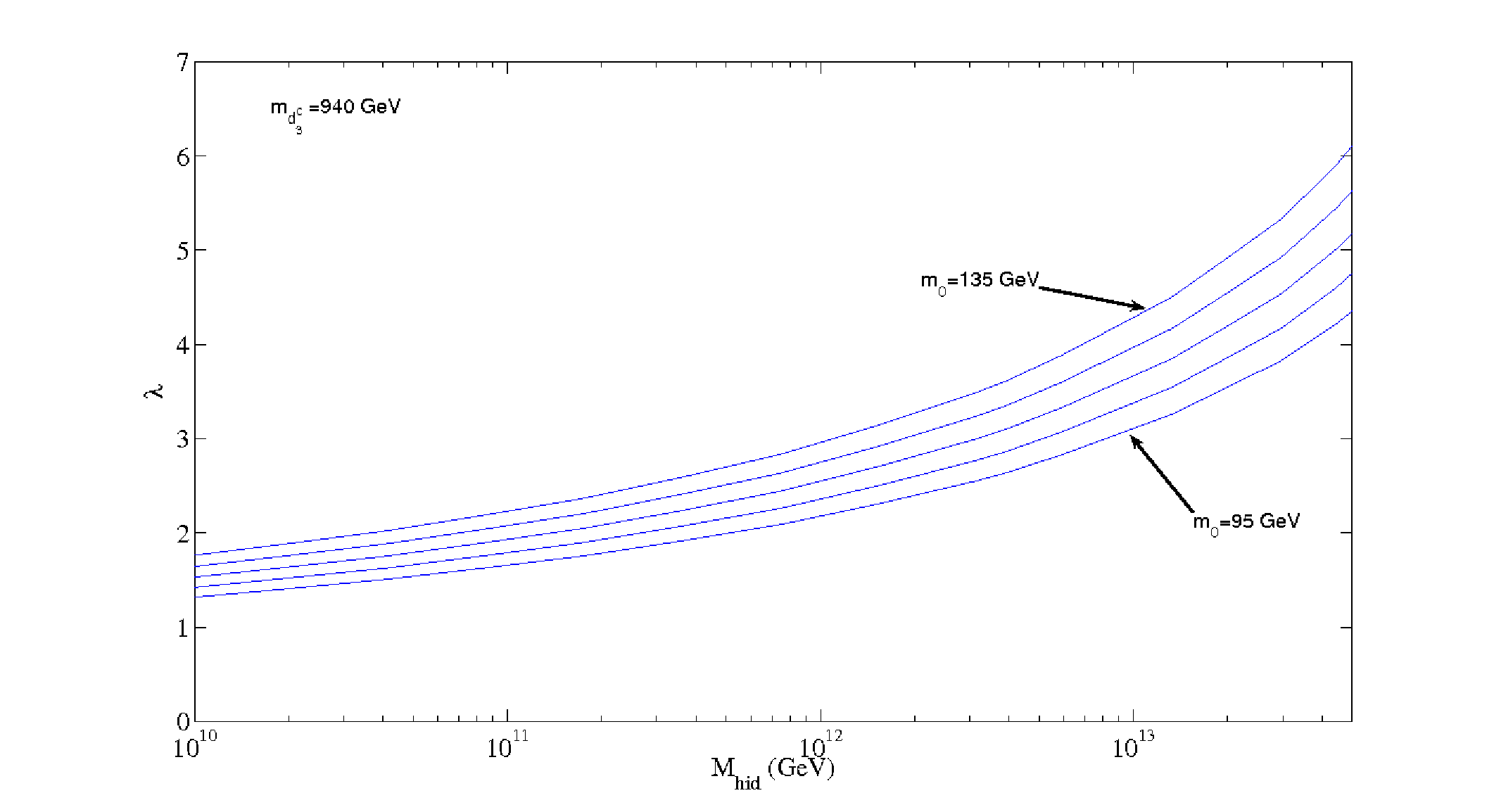}}
    \end{center}
    \caption{\it For a fixed low-energy prediction of a squark mass ($m_{d^c_3} = 940$~GeV in this case), 
    contours at different $m_0$ exist in the $M_{hid}$--$\lambda$ plane that predict the 
    same low-energy mass value.}
    \label{contour_indy_1}
 \end{figure}

However, we can now consider a different particle species and generate a similar set of contours 
within the same theoretical framework. By overlaying the two sets of contours and searching for the 
line of intersection consistent with our model assumption of universal scalar masses, we can 
reduce the parameter degeneracy of Fig.~\ref{contour_indy}. 
In Fig.~\ref{contour_indy_1}, we perform the same exercise as in Fig.~\ref{contour_indy},
but considering this time the third-generation right-handed down squark, 
assuming $m_{d^c_3} = 940$~GeV at the weak scale. In principle, we could take any 
species different from $L_3$, but by choosing the species most widely separated in gauge charges, 
we have a larger lever arm for obtaining the curve of intersection from the contour overlay. 

\begin{figure}[ht!]
    \newlength{\picwidthf}
    \setlength{\picwidthf}{6.2in}
    \begin{center}
        \resizebox{\picwidthf}{!}{\includegraphics{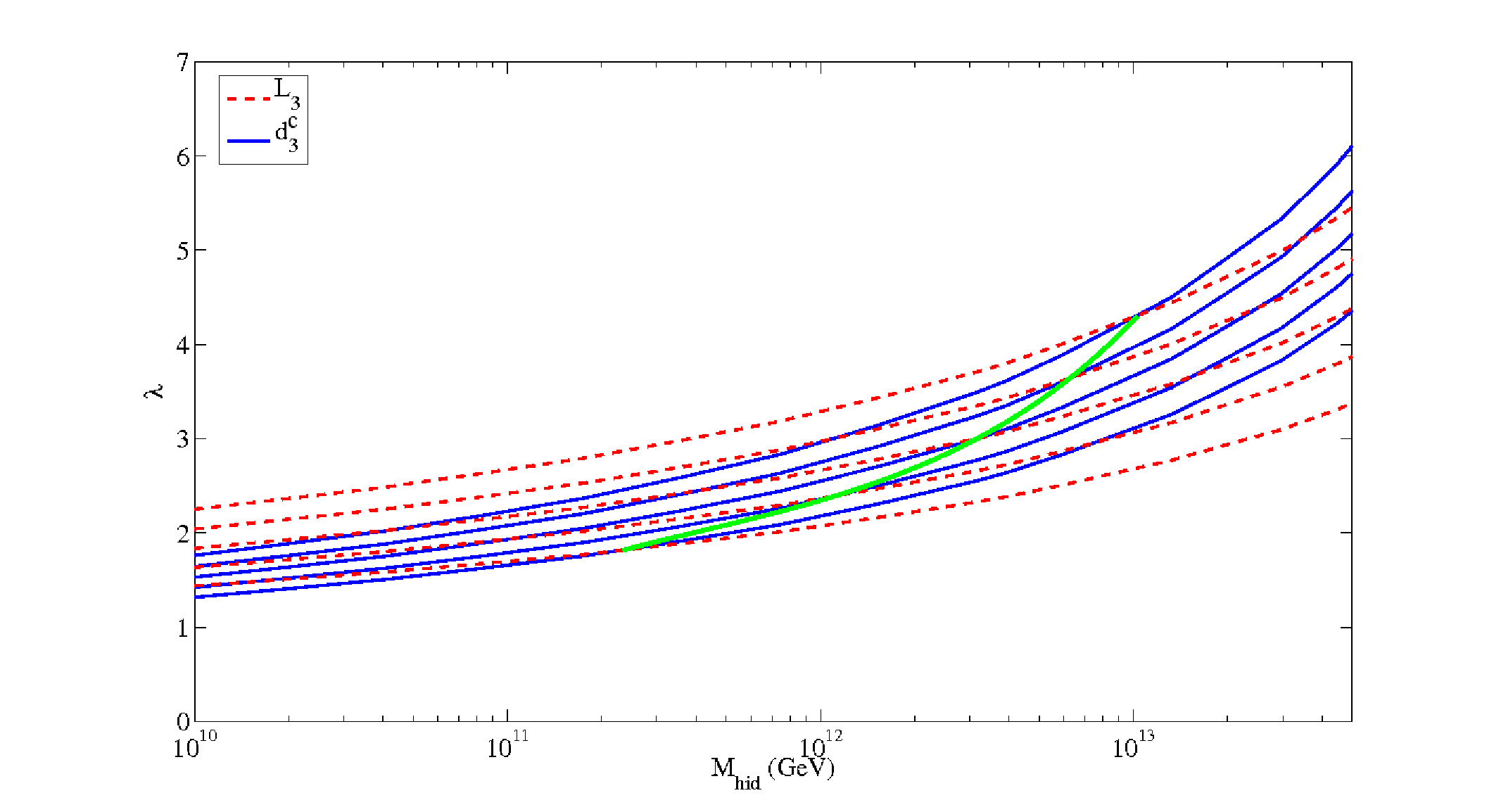}}
    \end{center}
 	\caption{\it Overlay of the contours in the $M_{hid}$--$\lambda$ plane of the 
	largest-eigenvalue left-handed slepton from Fig.~\protect\ref{contour_indy} (dashed lines) 
	with those of the largest-eigenvalue  right-handed squark from 
	Fig.~\protect\ref{contour_indy_1}(dark solid lines), assuming a universal soft 
	supersymmetry-breaking scalar mass $m_0$. The intersections (linked by a lighter
	solid line) show how measurements of the two masses can be combined to determine
	the values of $M_{hid}$ and $\lambda$.}
 	\label{contour_overlay_1}
 \end{figure}

We now take the contours of Fig.~\ref{contour_indy} and Fig.~\ref{contour_indy_1}, 
and overlay them in Fig.~\ref{contour_overlay_1}.
Under the assumption of universal scalar masses, the true theory must lie somewhere along the line of 
intersection indicated in the plot as a lighter (green) solid line. 
Whilst this line of intersection reduces the parameter space, we still do not have 
enough information to uncover uniquely the parameters of the hidden sector or the universal 
scalar mass. However, we can repeat the process of contour overlaying with two 
different species, and remove the parameter degeneracy. 

\begin{figure}[ht!]
    \newlength{\picwidthg}
    \setlength{\picwidthg}{6.2in}
    \begin{center}
        \resizebox{\picwidthg}{!}{\includegraphics{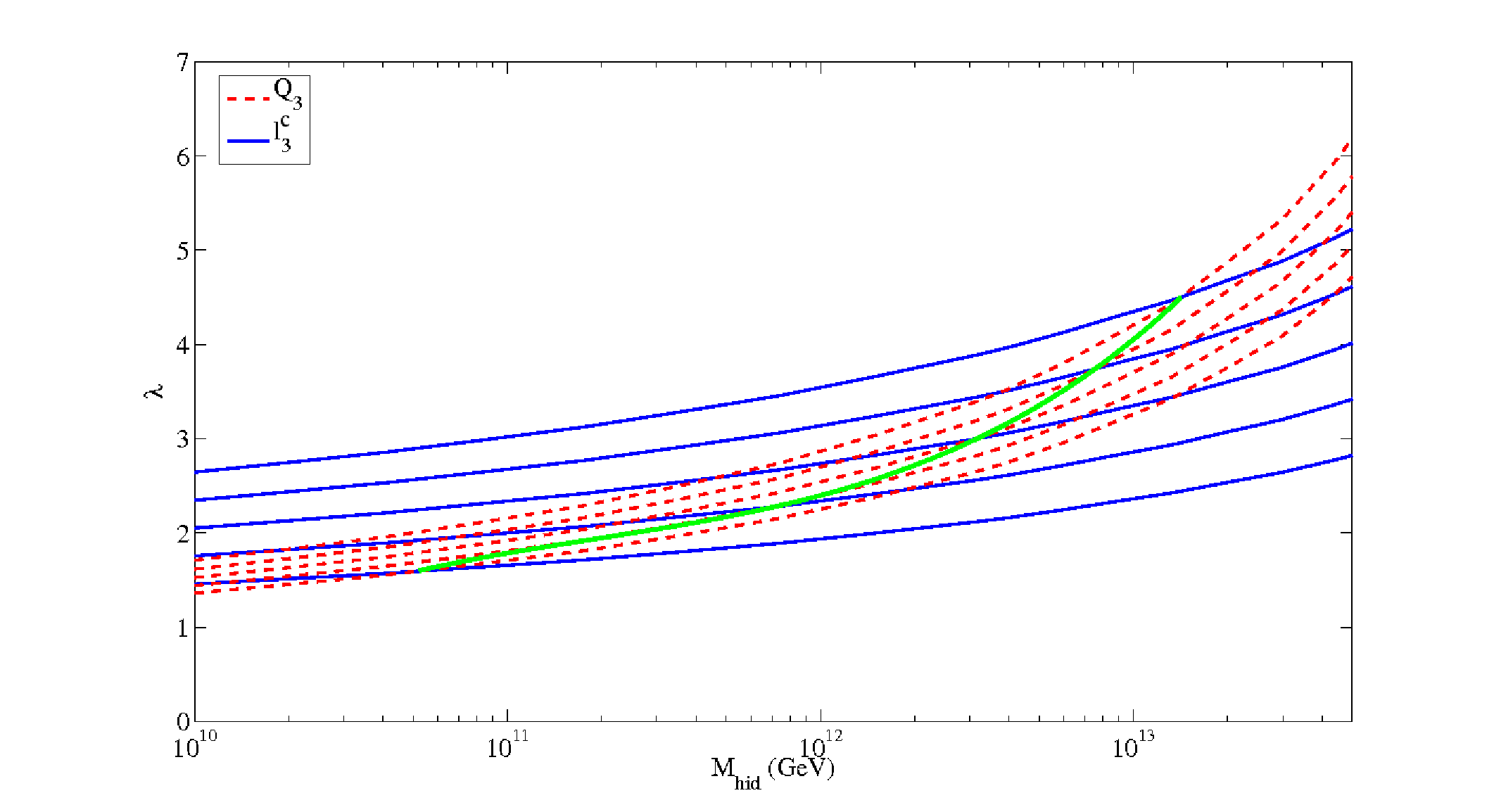}}
    \end{center}
 	\caption{\it As in Fig.~\protect\ref{contour_overlay_1}, but overlaying contours of the 
	largest-eigenvalue left-handed squark mass with contours of the 
	largest-eigenvalue right-handed slepton mass, again assuming a universal soft 
	supersymmetry-breaking 	scalar mass $m_0$.}
 	\label{contour_overlay_2}
 \end{figure}

In Fig.~\ref{contour_overlay_2} we superpose the third-generation 
left-handed squark, assuming $Q_3 = 940$~GeV at the weak scale, and the third-generation
right-handed slepton, assuming $l^c_3=160$~GeV at the weak scale.
We notice again a curve of intersection in Fig.~\ref{contour_overlay_2}, denoting consistency
with universal scalar masses. 

\begin{figure}[ht!]
    \newlength{\picwidthh}
    \setlength{\picwidthh}{6.2in}
    \begin{center}
        \resizebox{\picwidthh}{!}{\includegraphics{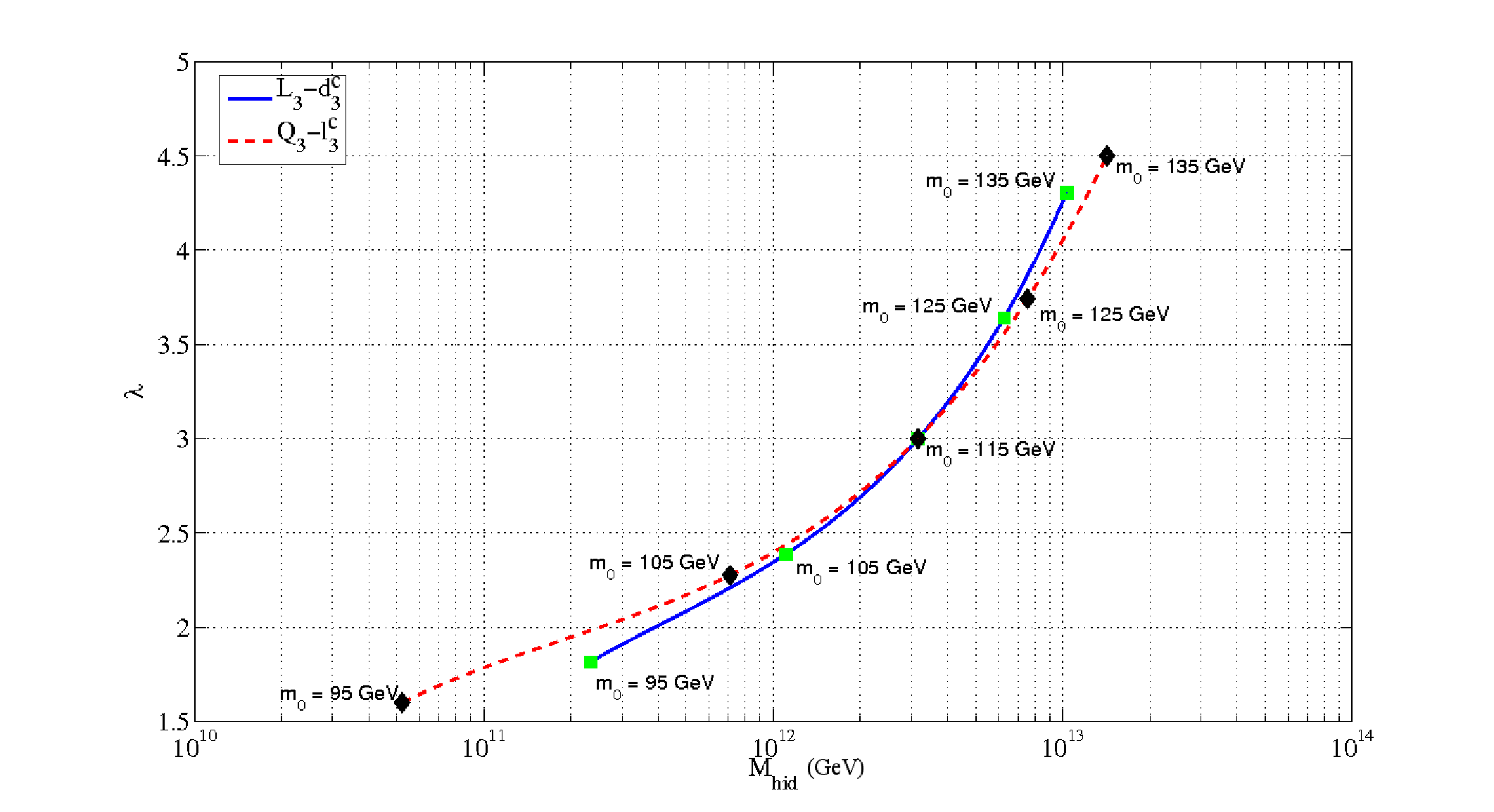}}
    \end{center}
 	\caption{\it The intersection of the intersection contours from 
	Figs.~\protect\ref{contour_overlay_1} and \protect\ref{contour_overlay_2}
	enables the parameters of the hidden sector to be
	reconstructed unambiguously from the low-energy data: the example chosen has 
	$m_0=115$~GeV, $\lambda = 3$, and $M_{hid} = 3.16\times 10^{12}$ GeV.}
 	\label{intersect_1}
 \end{figure}

We are now in a position to determine the parameters of the theory 
from the low-energy inputs. Having identified the curves of intersection of the contours using 
two different pairs of species (in this case $L_3$-$d^c_3$ and $Q_3$-$l^c_3$), we can now overlay 
the curves of intersection themselves. The crossing point of the two curves of intersection yields 
the unique point which determines both parameters of the hidden sector, $M_{hid}$ and 
$\lambda$, along with the universal scalar mass $m_0$. We can see the result for this example in 
Fig.~\ref{intersect_1}.

\begin{figure}[ht!]
    \newlength{\picwidthi}
    \setlength{\picwidthi}{6.2in}
    \begin{center}
        \resizebox{\picwidthi}{!}{\includegraphics{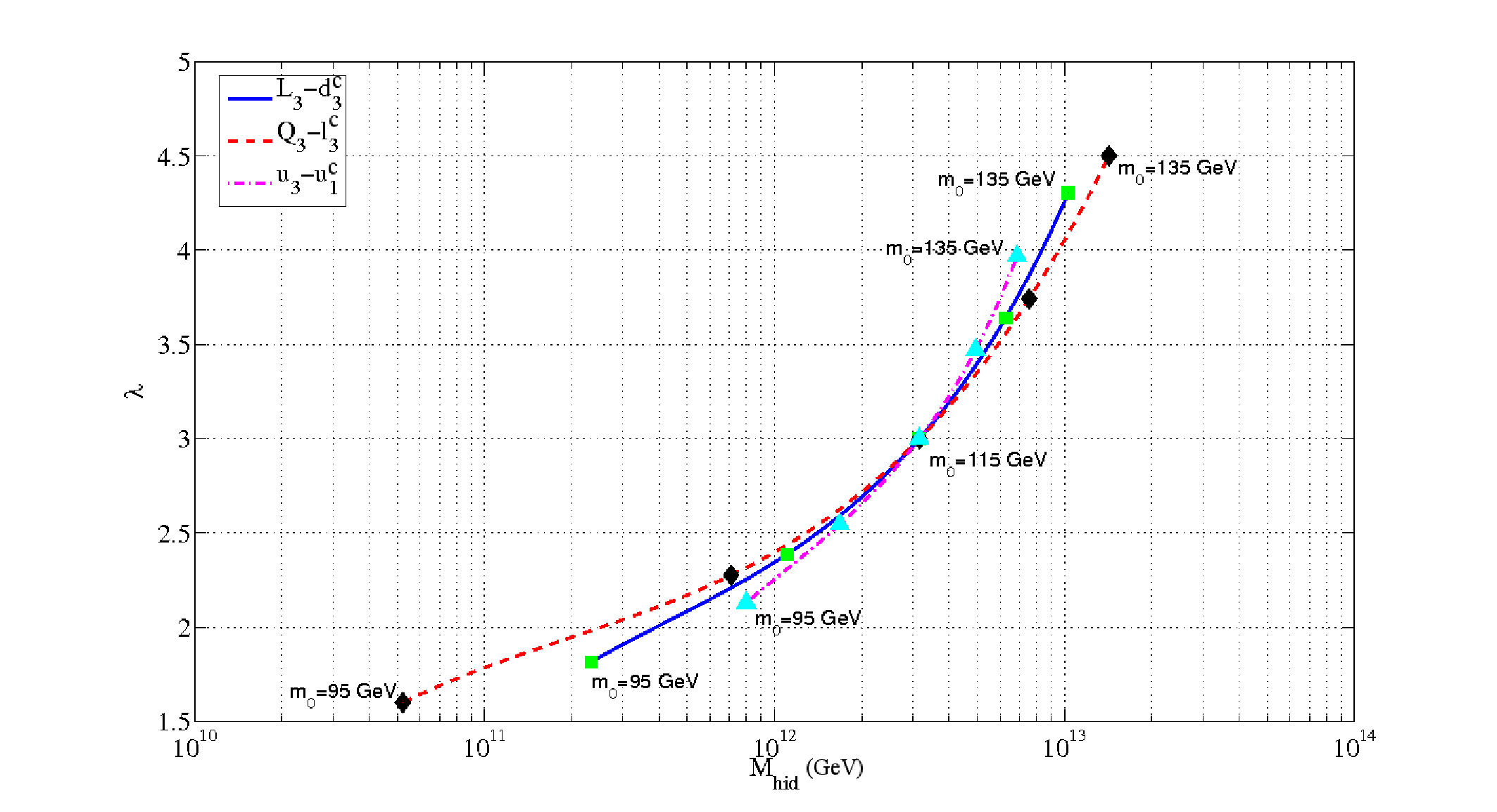}}
    \end{center}
 	\caption{\it A consistency check on the contour intersection shown in 
	Fig.~\protect\ref{intersect_1} is possible using a third set of low-energy data.
	In this example, $m_0=115$~GeV, $\lambda =3$, and $M_{hid} = 3.16\times 10^{12}$ GeV.}
 	\label{intersect_2}
 \end{figure}

We note that the universal scalar mass behaves like an affine parameter along each intersection 
curve. A consistent theory of the hidden sector with the assumption of universal scalar masses not 
only requires that the intersection of the contour-intersection curves, but also that the intersection 
occurs at the same affine parameter -- the universal scalar mass. As a consistency check on 
the method, we can repeat the process with yet another pair of species. 
Fig.~\ref{intersect_2} demonstrates the consistency check by repeating the process with the 
pair $u^c_3$-$u^c_1$ and adding the third contour-intersection curve to figure \ref{intersect_1}.

Whilst the method demonstrated in Figs.~\ref{contour_indy} to \ref{intersect_2} allowed us to 
reconstruct the hidden-sector parameters and the visible sector $m_0$, we required certain 
model assumptions. First, we assumed that the visible sector contained totally universal scalar 
masses and secondly that the hidden sector could be described by a two-parameter model. 
Under these mild assumptions we showed that we could recover a consistent model. If 
the reconstruction outlined above failed to recover a unique point in a plot similar to
Fig.~\ref{intersect_2}, then we would know that at least one of the model assumptions was incorrect.

In principle there are fifteen different observable eigenvalues in the slepton/squark sector: 
$L_{1,2,3}$; $l^c_{1,2,3}$; $Q_{1,2,3}$; $u^c_{1,2,3}$; $d^c_{1,2,3}$. Only the third-generation 
Yukawa couplings are large enough to cause significant generational splitting, which reduces 
the observable list to seven effective eigenvalues: $L_{(12)}$; $L_{3}$; $l^c_{(12)}$; $l^c_{3}$; 
$Q_{(12)}$; $Q_{3}$; $u^c_{(12)}$; $u^c_{3}$; $d^c_{(12)}$; $d^c_{3}$. The lightest eigenvalue in the 
up-like squark sector tracks the effect of the top Yukawa coupling, and if $\tan\beta$ is large 
enough for the bottom and tau Yukawa couplings to be usefully large they will do likewise. 
With the 21 different possible pairings over-constraining the system, we can test model 
consistency, as illustrated in our example above.

On the other hand, the method outlined above is not restricted to models with total universality 
among the scalar masses. Results showing the absence of exotic contributions of 
flavour-changing neutral-current (FCNC) interactions in the kaon and B-meson system require 
a high level of mass degeneracy ($\Delta m \lesssim 10^{-4}$) between scalar particles carrying 
the same gauge quantum numbers -- the super-GIM mechanism. Even if we assume the least 
amount of universality consistent within the model class we consider by requiring all scalar 
particles with the same gauge quantum numbers appear with a common mass at $M_X$, 
we can still reconstruction the model parameters of the hidden sector. In this case, we can use the 
large third-generation Yukawas, that create generational splitting, to provide a lever arm for
reconstructing the hidden-sector parameters. We demonstrate this reconstruction in our second example.

\subsection{Example 2: Non-Degenerate Scalars and the Top Yukawa Coupling}

\begin{figure}[ht!]
    \newlength{\picwidthj}
    \setlength{\picwidthj}{6.2in}
    \begin{center}
        \resizebox{\picwidthj}{!}{\includegraphics{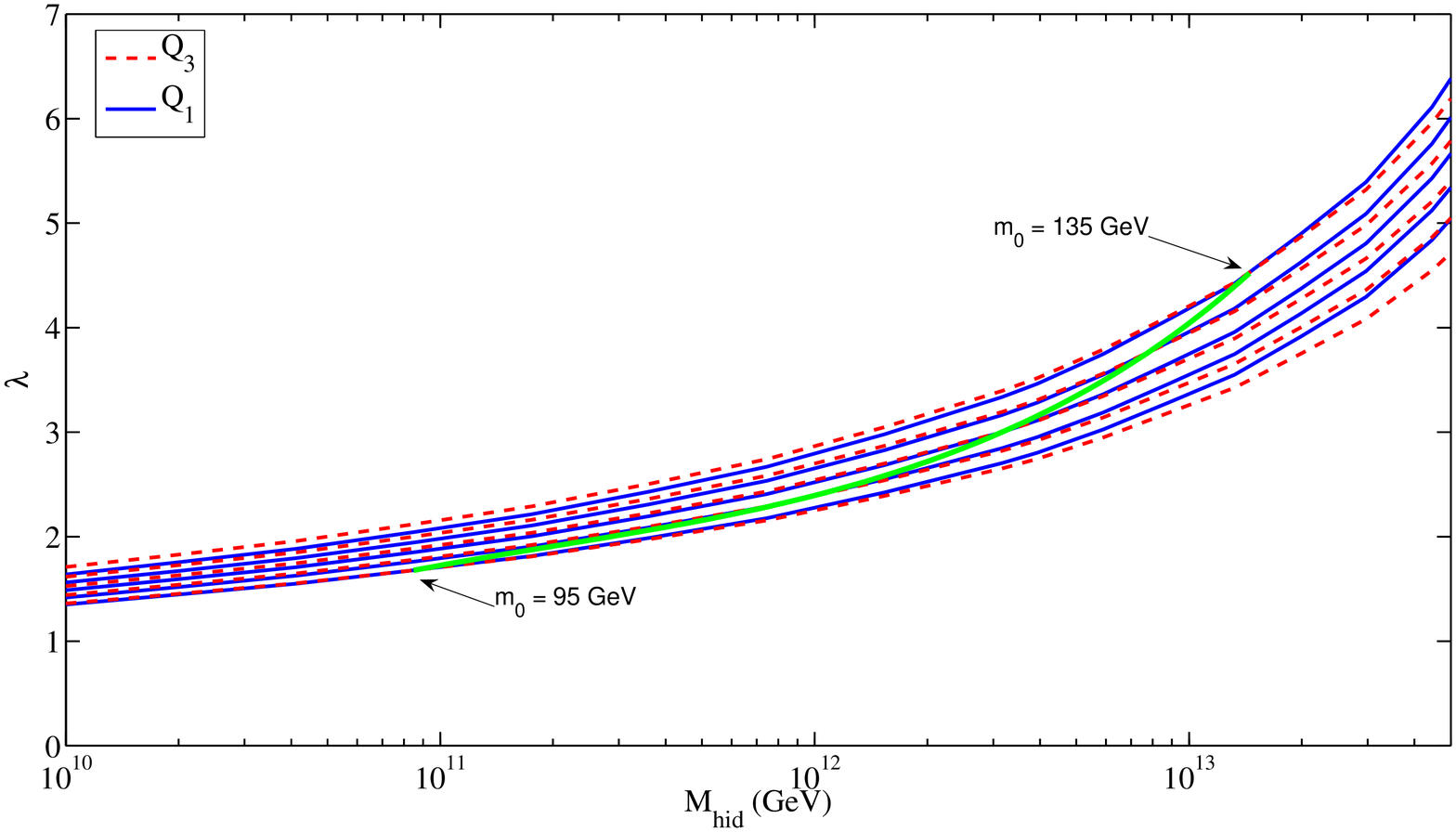}}
    \end{center}
 	\caption{\it Overlay of the contours of $Q_1$ and $Q_3$ masses, assuming only 
	that they have a common scalar mass at the input scale.}
 	\label{Q-Q}
 \end{figure}

\begin{figure}[ht!]
    \newlength{\picwidthk}
    \setlength{\picwidthk}{6.2in}
    \begin{center}
        \resizebox{\picwidthk}{!}{\includegraphics{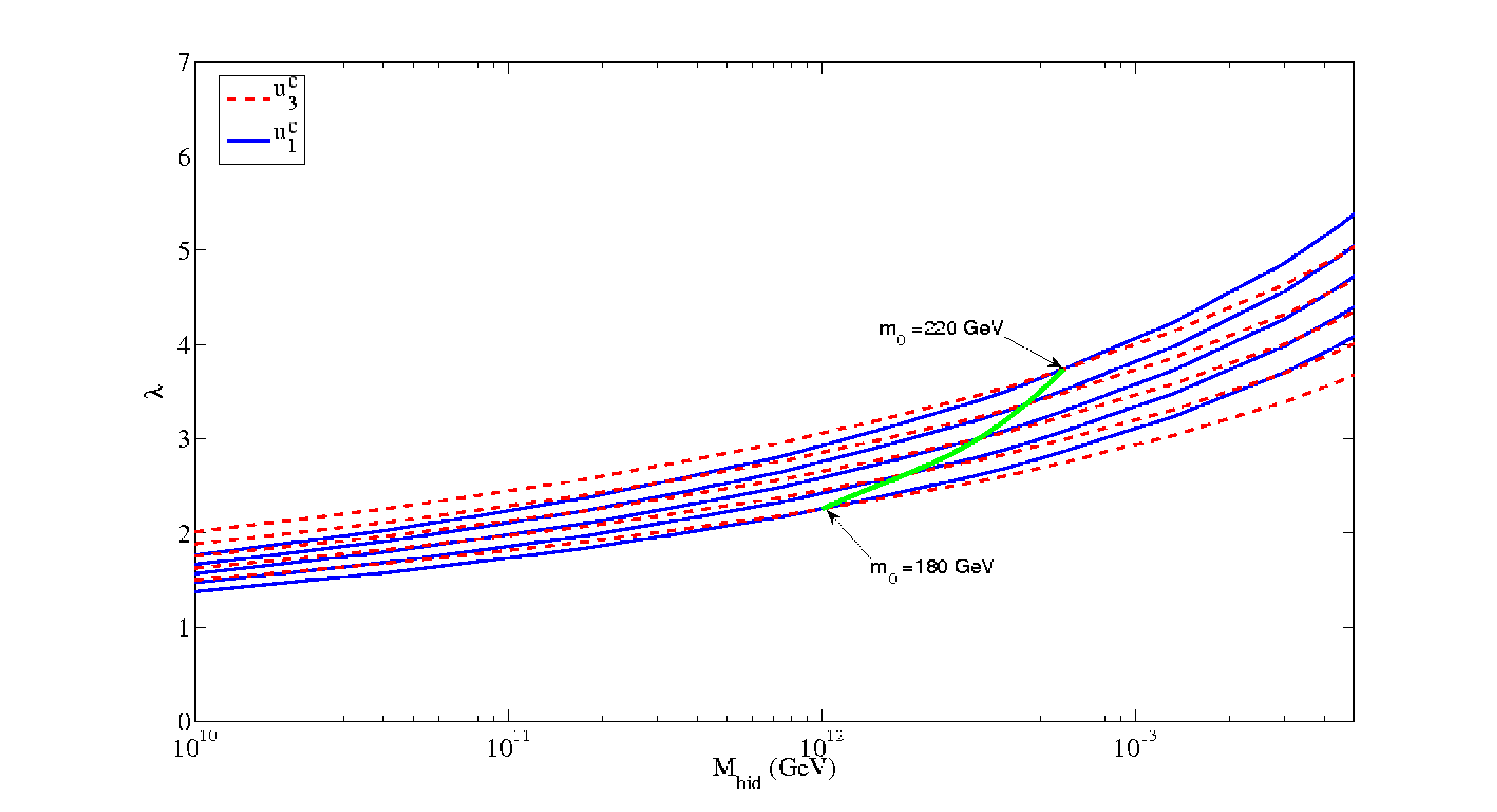}}
    \end{center}
 	\caption{\it Overlay of the contours of the $u^c_1$ and $u^c_3$ masses, assuming only 
	that they have a common scalar mass at the input scale.}
 	\label{u-u}
 \end{figure}

We again assume a model of the hidden sector as given in eq.(\ref{Inter}). However,
in this example we assume a common mass only for squarks with the same gauge quantum numbers, namely $Q_{i}(M_X)=m_0$ and $u^c_{i}(M_X)=m_0^\prime \ne m_0$ in general. 
We then proceed analogously to the previous example, this time generating contours 
in the $M_{hid}$--$\lambda$ plane for the $Q_1-Q_3$ and $u^c_3-u^c_1$ systems. 
The differences between the contours are due to the top-quark Yukawa contributions to the RGEs
for the third-generation squarks. In Fig.~\ref{Q-Q} we overlay the contours corresponding to 
different input soft supersymmetry-breaking masses for $Q_3-Q_1$, and
the lighter-coloured (green) solid line intersecting the contours corresponds to common values for
the input mass. Repeating this process with the $u^c_3-u^c_1$ system yields Fig.~\ref{u-u}.

\begin{figure}[ht!]
    \newlength{\picwidthl}
    \setlength{\picwidthl}{6.2in}
    \begin{center}
        \resizebox{\picwidthl}{!}{\includegraphics{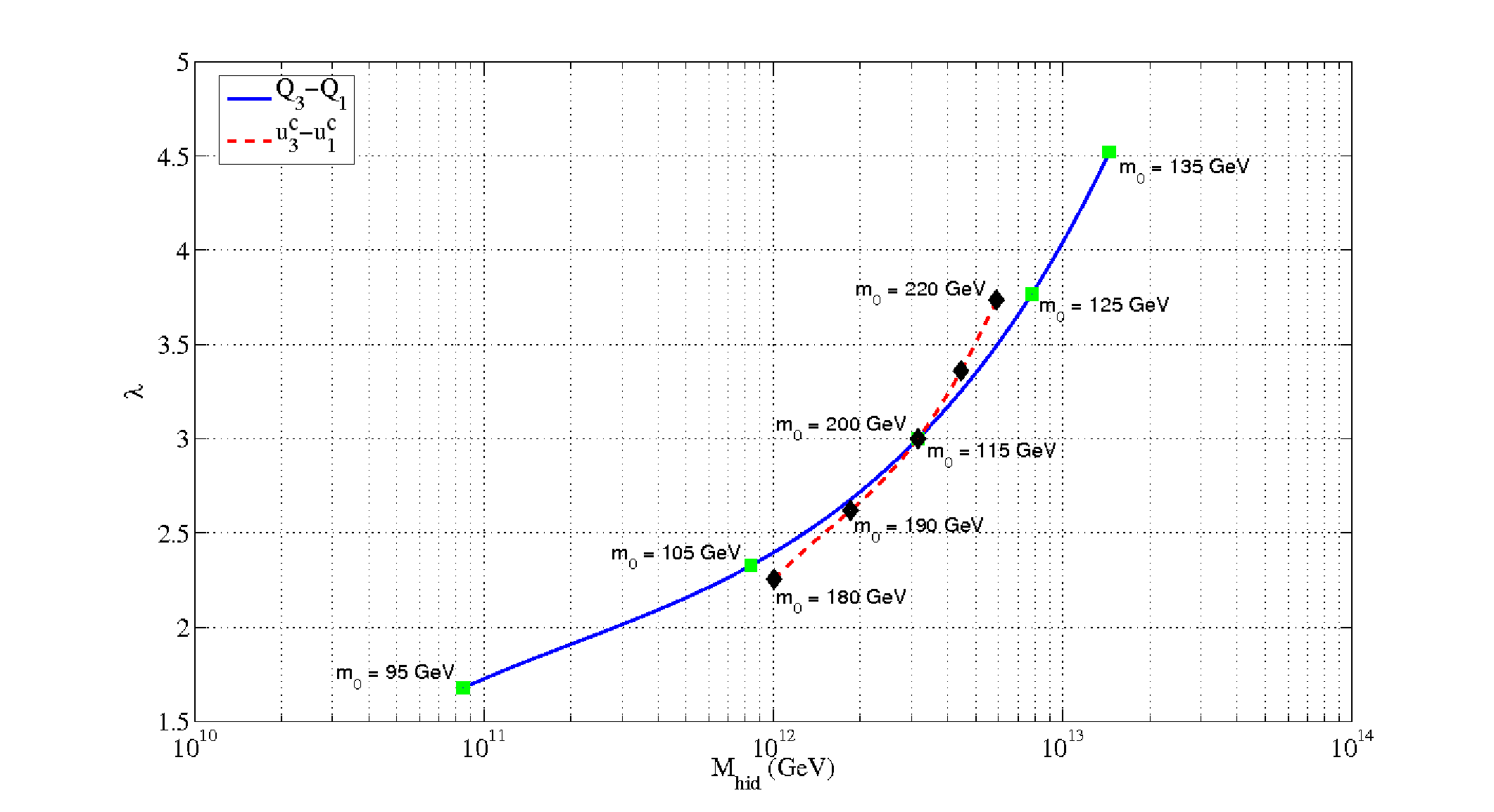}}
    \end{center}
 	\caption{\it Intersection of the intersection curves from the previous two figures, yielding
	consistent parameters of the hidden sector reconstructed from the low-energy data.
	We allow $m_0^\prime \ne m_0$ in this example.}
 	\label{common_scalar}
 \end{figure}

At this point, we overlay the two curves of intersection from Figs.~\ref{Q-Q} and \ref{u-u},
and examine the crossing point as shown in Fig.~\ref{common_scalar}. We recall that we
allow $m_0^\prime \ne m_0$ in this example. We again
see that we can extract the parameters of the hidden sector via the intersection point in the 
$M_{hid}$ -- $\lambda$ plane. We recall that the visible sector parameter $m_0$ serves as an 
affine parameter along each curve. Notice that in this case we obtain two different values for the 
visible sector parameter $m_0$, since the curves intersect at different values of their respective 
affine parameters. Thus, even in the more conservative case in which we assume a common 
scalar masses only among particles with the same gauge quantum numbers, consistent with a
super-GIM mechanism, we can still reconstruct the parameters of the hidden sector by concentrating 
on the effects of the large Yukawa couplings.

With only  the $Q_1-Q_3$ and $u^c_1-u^c_3$ systems, we would not be able to  discriminate 
between different models of the hidden sector. However, for sufficiently large $\tan\beta$, 
the third-generation down quark also has a large Yukawa coupling (and if $\tan \beta$ is large
enough we may be able to repeat the procedure with a large $\tau$ Yukawa coupling). 
Then we can use the $Q_1-Q_3$ and $d^c_1-d^c_3$ system, to get an independent determination 
of the hidden-sector parameters. In general, these two determinations of the hidden-sector 
parameters will agree only if we are reconstructing correctly the model parametrization of the 
hidden-sector dynamics. As such, with the two independent determinations, we can determine 
both the correct parametrization of the hidden sector, and fit its parameters with a redundant check. 

\subsection{General Reconstruction Strategy}

These two examples demonstate a general strategy that can be employed with the low-energy 
data to determine model parameters of the hidden sector, which we summarize as follows: 
\begin{itemize}
	\item{Using the measured low-energy data on the soft supersymmetry-breaking
scalar masses, run the MSSM RGEs bottom-up and check for the appearance of a mirage 
scalar unification scale, distinct from the gauge coupling unification scale, $M_X$. Mirage 
scalar unification would signify the potential presence of a hidden-sector effect.}
	\item{Starting with the contours in the $Q_3-Q_1$ and $u^c_3-u^c_1$ systems constructed, 
as described above, from a model of the hidden sector together with the conservative assumption 
of common scalar masses among particles with the same gauge quantum numbers, construct the 
curves of intersection and overlay them in the $M_{hid}$--$\lambda$ plane. Using the $Q_3-Q_1$ 
and $u^c_3-u^c_1$ system takes advantage of the large top Yukawa coupling, which provides the 
largest generational splitting in the RGE flow.}
        \item{If $\tan\beta$ is sufficiently large (which in practice is determined by the following 
procedure yielding a definite result), starting with contours in the $Q_3-Q_1$ and $d^c_3-d^c_1$ 
systems constructed from a model of the hidden sector together with the conservative assumption 
of common scalar masses among particles with the same gauge quantum numbers, construct the 
curves of intersection and overlay them in the $M_{hid}$--$\lambda$ plane. Using the $Q_3-Q_1$ 
and $d^c_3-d^c_1$ system takes advantage of the potentially large (enhanced at $\tan\beta$)
bottom Yukawa coupling, which provides the second-largest generational splitting in the RGE
flow. Try both this step and the previous one for each of the different hidden-sector parametrizations; 
the hidden-sector parametrization which most correctly repesents the qualitative behaviour of the 
hidden sector renormalization effects will be the one for which the two parameter determinations 
best agree. This then determines both the correct parametrization for the hidden sector and gives 
a redundant determination of the parameters.} 
	\item{For the intersections established in the previous steps, the affine parameter along 
each curve will yield the common scalar mass for each system at the intersection point, 
but these could be different for the two curves in the intersections of lines of intersection. If the 
extracted common scalar masses are the same for both systems (in each of the 
reconstructions proposed above if both are feasible), this would indicate that all scalar particles 
may have universal masses.}
	\item{If the $Q_3-Q_1$ and $u^c_3-u^c_1$ system yields the same common scalar mass 
at intersection (and the $Q_3-Q_1$ and $d^c_3-d^c_1$ system does likewise, assuming that 
this reconstruction is feasible) apply the universality assumption using the parameters determined 
from the $Q_3-Q_1$ and $u^c_3-u^c_1$ system (and  $Q_3-Q_1$ and $d^c_3-d^c_1$ system if it is
feasible and in agreement) and predict $l_{(12)}$; $l^c_{3}$~\footnote{We note that $L_{(12)}$; 
$L_{3}$ are in general affected by renormalization effects associated with the seesaw interactions, 
as are $N_{(12)}$; $N_{3}$; we discuss hidden-sector effects and the seesaw mechanism in 
a companion paper.}. In the case of a successful prediction, we would have a completely 
self-consistent model of the hidden sector with knowledge of the boundary conditions for the 
scalar masses at the true unification scale. If the prediction does not yield a successful prediction, 
then either the assumption on universality is incorrect or potentially the model of the hidden sector 
is wrong, in the case that one was not able to do the redundant check using the $Q_3-Q_1$ and 
$d^c_3-d^c_1$ system. (In this case, we can repeat the process with a different model of the hidden
sector.) }
\end{itemize}

%%%%%%%%%%%%%%%%%%%%%%%%%%%%%%%%%%%%%%%%%%%%%%%%%%%%%%%
%%%%%%%%%%%%%%%%%%%%%%%%%%%%%%%%%%%%%%%%%%%%%%%%%%%%%%%
%%%%%%%%%%%%%%%%%%%%%%%%%%%%%%%%%%%%%%%%%%%%%%%%%%%%%%%
%%%%%%%%%%%%%%%%%%%%%%%%%%%%%%%%%%%%%%%%%%%%%%%%%%%%%%%

\section{Reconstruction Algorithm}

We now review the steps that would be needed, starting from experimental measurements, 
to characterize and parametrize the hidden sector, following the approach proposed in the 
previous section. 

Clearly all of the above considerations are moot if the LHC does not actually discover supersymmetry. 
However, it is important not only that the LHC finds some evidence for supersymmetry, 
in the form of certain superpartners, but also that the LHC measures the masses of sufficiently many of
the spartners of Standard Model particles to be able to make redundant consistency checks. It is 
also important to know whether any other chiral multiplets carrying Standard Model charges
appear below the GUT scale. Our analysis above was dependent on the RGE running of the
MSSM parameters between the high scales of mediation and the hidden sector on one hand,
and the electroweak scale on the other hand, and we assumed there is a large desert in between. 
To be able to extrapolate reliably
through the desert, we need to know the entire matter content of the effective theory at desert 
energies. 

In particular, to use our algorithm we need to assume that the LHC (if necessary in
combination with a linear $e^+ e^-$ collider) is able to determine the complete superspectrum 
below the desert. This is a necessary input into any determination of gauge-coupling unification 
in supersymmetric extensions of the Standard Model. The MSSM has the feature that RGE 
extrapolation based on its particle content yields gauge-coupling unification at a unification 
scale of order $10^{16}$ GeV. With the LHC, we trust that we will no longer be assuming some 
low-energy field particle content, but rather we will be in a position to determine it from experiment. 
To get the extrapolation right, and really be able to test gauge-coupling unification, we need to 
know the complete list of supermultiplets which are dynamical at desert energies. In addition,
one would also like the reduce the (controllable) theory uncertainties as much as possible: 
these include, in particular, higher-order terms in the $\beta$ functions for the RGE running, 
and the TeV-scale threshold corrections for the observed spectrum of input 
masses, which should also be incorporated at high order.

If, after the complete spectrum of superpartners and their physical masses has been determined, 
one finds that the gauge-coupling extrapolation does indeed lead to unification at a large  
scale of ${\cal O}(10^{16})$~GeV, one can then ask about the gauginos and scalars.
As we have discussed above, gauginos receive mass renormalization from the hidden-sector 
only through external-line wave function renormalization of the singlet superfield whose 
F-term is responsible for supersymmetry breaking. If we adopt a renormalization prescription 
in which this wave-function renormalization is absorbed into the VEV, then the effects of 
the renormalization on all the soft masses can be absorbed into a common rescaling of the 
boundary value at the mediation scale, and in this renormalization scheme the gauginos are 
not directly affected by the hidden sector. 

This means that the gauginos give a clean 
characterization of the high-scale mechanisms of supersymmetry mediation and breaking, 
unobscured by hidden-sector effects. The generality and simplicity of the patterns of gaugino 
masses that may arise in modern supersymmetric unified theories has been emphasized in the 
review~\cite{cn:07}. By comparing the gaugino mass pattern observed at the LHC (and
possibly a linear collider) to these theoretical expectations, one may get an indication of the 
model classes preferred by the data. This would in turn give some indication of the expectations 
we may have for the high-scale scalar masses input at the mediation or unificaition scale.

In order to try to use hidden-sector effects on the scalar masses to observe the dynamics of the 
hidden sector, a key question will be the degree of scalar-mass universality. Universality of
the sfermions with identical Standard Model gauge charges seems very plausible~\cite{EN,BG,il:02,cam:02}, but 
will need to be checked. Many more consistency checks would be possible if there is a higher
degree of scalar-mass universality, as in certain GUTs or the CMSSM. In order to check these
hypotheses, one will need to measure the mass-squared parameters for the scalars of different 
gauge charges, and separately for the squarks and sleptons of the third generation. Doing this 
precisely will presumably necessitate studying the scalars not only at the LHC, but also at a linear
$e^{+}e^{-}$ collider, such as CLIC, capable of producing all the scalars of the 
supersymmetric extension of the Standard Model, especially including the squarks which are the 
starting point of our reconstruction algorithm. 

The precision of an $e^{+}e^{-}$ collider will be 
important, as the scalar masses-squared that we use in our reconstruction are only parts of entries 
in the mass-squared matrix for the chiral scalars that are the partners of the left- and right-handed
quarks or leptons. One will need to fix the other entries in the mass-squared matrix from data, 
and that will necessitate fixing several of the soft supersymmetry-breaking parameters, 
including A-terms. It will also be necessary to determine $\tan \beta$ quite accurately.
The required level of detail and precision will be hard to achieve without combining 
LHC studies with follow-up precision studies at an $e^{+}e^{-}$ collider. 

We note that the reconstruction procedure outlined in the previous section was exemplified on the 
basis of scalars with a general pattern we may characterize as gravity/modulus mediated. That 
was not essential to the reconstruction; mirage/anomaly mediation patterns of high-scale input 
scalar masses would have been equally adapted to our arguments. On can use the gaugino 
mass patterns from the previous paragraph to infer the type of unification theory, and hence the 
type of input pattern to consider. It will be checked {\it a posteriori} by success in finding a 
consistent set of hidden-sector parameters that matches the input pattern of scalar masses
within the class of models. 

Also, with a given pattern of input scalar masses (e.g., universal as in pure gravity mediation) there 
will be linear combinations of the low-energy scalar mass parameters which are not affected by 
the hidden-sector renormalization (at leading order in the observable-sector 
couplings). These sum rules can be used to test whether the observed distortion of 
the scalar spectrum is due in fact to hidden-sector effects and not some other new high-scale 
physics. We enumerate these sum rules in Appendix A, for the case of universal scalar masses
(see also~\cite{kkm:08}). We note that, beyond leading order, RGE feedback from observable-sector
interactions will spread the hidden-sector effects through all the soft parameters, so that
complete simultaneous fits to the soft parameters (gaugino masses included) would be essential.

%%%%%%%%%%%%%%%%%%%%%%%%%%%%%%%%%%%%%%%%%%%%%%%%%%%%%%%
%%%%%%%%%%%%%%%%%%%%%%%%%%%%%%%%%%%%%%%%%%%%%%%%%%%%%%%
%%%%%%%%%%%%%%%%%%%%%%%%%%%%%%%%%%%%%%%%%%%%%%%%%%%%%%%
%%%%%%%%%%%%%%%%%%%%%%%%%%%%%%%%%%%%%%%%%%%%%%%%%%%%%%%

\section{Conclusions}

We have demonstrated in this paper that it is possible, in principle, to extract
dynamical parameters of the hidden sector responsible for supersymmetry breaking
from measurements of sparticle masses in the observable sector. Hidden-sector
dynamics could affect the renormalization of soft supersymmetry-breaking scalar
masses between the input scale $M_X$ and the hidden-sector scale $M_{Hid}$.
This extra renormalization would make it seem as if observable squark and slepton
masses would unify at some `mirage' scale below $M_X$, if only the MSSM RGEs
were used to run upwards. We showed how the parameters of the hidden sector
could be measured using low-energy data, and consistency checks performed.
We demonstrated the procedure in two examples, one assuming
total universality of the soft supersymmetry-breaking scalar masses, and one
assuming universality only for sfermions with the same gauge quantum numbers,
and gave a general prescription for extracting dynamical parameters of the hidden sector.

Several aspects of this programme require further study. It is unclear whether LHC
measurements by themselves would provide enough low-energy information,
and we anticipate that information from a linear collider would also be required:
its precision and ability to determine slepton mass parameters would both be
important for our programme. We have not quantified the accuracy with which
soft supersymmetry-breaking masses should be measured in order to obtain
intersting accuracy in the hidden-sector parameters. This might actually be
premature in the absence of a credible model of the hidden sector to use as a
benchmark. Also, we have not included higher-order terms in the RGEs for the
soft supersymmetry-breaking parameters: this is important to do, though we do not expect
our essential conclusions to be altered significantly. Finally, we note that it would be 
interesting to extend the approach described here to lower-scale models of
supersymmetry breaking, such as gauge mediation.

We hope that this work opens the way to further studies of the possibility of measuring 
observable effects of the hidden-sector dynamics. The basic motivation for such a programme
is clear. There is certainly no shortage of scope for further studies, and in a companion
paper~\cite{cem:2} we explore some possible implications of the hidden sector for
flavour studies, particularly in the lepton sector. Perhaps the concept of hidden-sector
phenomenology is not an oxymoron.

%%%%%%%%%%%%%%%%%%%%%%%%%%%%%%%%%%%%%%%%%%%%%%%%%%%%%%%
%%%%%%%%%%%%%%%%%%%%%%%%%%%%%%%%%%%%%%%%%%%%%%%%%%%%%%%
%%%%%%%%%%%%%%%%%%%%%%%%%%%%%%%%%%%%%%%%%%%%%%%%%%%%%%%
%%%%%%%%%%%%%%%%%%%%%%%%%%%%%%%%%%%%%%%%%%%%%%%%%%%%%%%

\section*{Acknowledgments}

We are deeply grateful to Graham Ross for sharing with us the parametrization of hidden sector renormalization effects which 
we have employed in Section 3 of this work, 
as well as for helpful discussions and encouragement. 
We would like to thank John March-Russell and Stephen West 
for useful discussions. We would also like to acknowledge the 
support of the Natural Sciences and Engineering Research 
Council of Canada.

%%%%%%%%%%%%%%%%%%%%%%%%%%%%%%%%%%%%%%%%%%%%%%%%%%%%%%%
%%%%%%%%%%%%%%%%%%%%%%%%%%%%%%%%%%%%%%%%%%%%%%%%%%%%%%%
%%%%%%%%%%%%%%%%%%%%%%%%%%%%%%%%%%%%%%%%%%%%%%%%%%%%%%%
%%%%%%%%%%%%%%%%%%%%%%%%%%%%%%%%%%%%%%%%%%%%%%%%%%%%%%%

\section*{Appendix A: Scalar-Mass Sum Rules in the Presence of a Hidden Sector}

%%%%%%%%%%%%%%%%%%%%%%%%%%%%%%%%%%%%%%%%%%%%%%%%%%%%%%%
%%%%%%%%%%%%%%%%%%%%%%%%%%%%%%%%%%%%%%%%%%%%%%%%%%%%%%%

When hidden-sector renormalization of the scalar mass operator is combined with the 
observable-sector gauge and Yukawa interactions, there is a distortion of the mass pattern of the 
scalars at low energies . However, at one-loop order in the visible-sector couplings, and to all 
orders in the hidden-sector couplings, some relations are preserved~\cite{crs:07,kkm:08}. 
For the case of universal input scalar masses, without further assumptions one has:
%%%%%%%%%%%%%%%%%%%%%%%%%%%%%%%%%%%%%%%%%%%%%%%%
\begin{equation}
  \label{eq:sr1}
   m_{\twi Q}^2-2 m_{\twi u^c}^2 + m_{\twi d^c}^2 -m_{\twi L}^2 +
   m_{\twi e^c}^2 = 0
\end{equation}
%%%%%%%%%%%%%%%%%%%%%%%%%%%%%%%%%%%%%%%%%%%%%%%%
\begin{equation}
\label{eq:sr18}
  2 \smass{Q_{3}} -\smass{u^c_{3}}-\smass{d^c_{3}} -2 \smass{Q_{1}}
  +\smass{u^c_{1}}+\smass{d^c_{1}} = 0
\end{equation}
%%%%%%%%%%%%%%%%%%%%%%%%%%%%%%%%%%%%%%%%%%%%%%%%
If $\tan\beta$ is small one also has:
%%%%%%%%%%%%%%%%%%%%%%%%%%%%%%%%%%%%%%%%%%%%%%%%
\begin{equation}
  \label{eq:sr20}
  2\smass{Q_{3}} - \smass{u^c_{3}}  - 2\smass{Q_{1}} + \smass{u^c_{1}}=
  0
\end{equation}
%%%%%%%%%%%%%%%%%%%%%%%%%%%%%%%%%%%%%%%%%%%%%%%
\begin{equation}
\label{eq:sr21}
  \smass{e^c_{3}} -  \smass{e^c_{1}} = 0
\end{equation}
%%%%%%%%%%%%%%%%%%%%%%%%%%%%%%%%%%%%%%%%%%%%%%%%
If there are no neutrino seesaw effects visible in the soft masses one also has:
%%%%%%%%%%%%%%%%%%%%%%%%%%%%%%%%%%%%%%%%%%%%%%%%
\begin{equation}
\label{eq:sr19}
  2\smass{L_{3}} - \smass{e^c_{3}} -2\smass{L_{1}} + \smass{e^c_{1}} = 0
\end{equation}
%%%%%%%%%%%%%%%%%%%%%%%%%%%%%%%%%%%%%%%%%%%%%%%%
If the $\mu$ term does not arise from a Giudice-Masiero mechanism and 
$m_{H} = m_{\bar{H}}$ one also has:
%%%%%%%%%%%%%%%%%%%%%%%%%%%%%%%%%%%%%%%%%%%%%%%%
\begin{equation}
\label{eq:sr22}
  3\smass{u^c_{3}}-3 \smass{d^c_{3}}+ 2\smass{L_{3}}-2 \smass{e^c_{3}}
  -2m^{2}_{H}+2 m^{2}_{\bar{H}} -3\smass{u^c_{1}}+3 \smass{d^c_{1}}-
  2\smass{L_{1}}+2 \smass{e^c_{1}}= 0
\end{equation}
%%%%%%%%%%%%%%%%%%%%%%%%%%%%%%%%%%%%%%%%%%%%%%%%%
If the $\mu$ term does not arise from a Giudice-Masiero mechanism and 
$m_{H} = m_{\bar{H}} = m_{\tilde{q}} = m_{\tilde{l}}$ one also has:
%%%%%%%%%%%%%%%%%%%%%%%%%%%%%%%%%%%%%%%%%%%%%%%%%
\begin{equation}
\label{eq:sr23}
  -3 \smass{d^c_{3}} -\smass{e^c_{3}} + 2 m^{2}_{\bar{H}} +3 \smass{d^c_{1}}
  -2\smass{L_{1}}+\smass{e^c_{1}} = 0
\end{equation}
%%%%%%%%%%%%%%%%%%%%%%%%%%%%%%%%%%%%%%%%%%%%%%%%%%
These relations allow one to check the consistency of various assumptions on the soft parameters 
with RGE evolution including contributions from the hidden sector. Note that they are only true to 
leading order in the observable-sector interactions. 

%%%%%%%%%%%%%%%%%%%%%%%%%%%%%%%%%%%%%%%%%%%%%%%%%%%%%%%
%%%%%%%%%%%%%%%%%%%%%%%%%%%%%%%%%%%%%%%%%%%%%%%%%%%%%%%
%%%%%%%%%%%%%%%%%%%%%%%%%%%%%%%%%%%%%%%%%%%%%%%%%%%%%%%
%%%%%%%%%%%%%%%%%%%%%%%%%%%%%%%%%%%%%%%%%%%%%%%%%%%%%%%
%%%%%%%%%%%%%%%%%%%%%%%%%%%%%%%%%%%%%%%%%%%%%%%%%%%%%%%

\section*{Appendix B: MSSM Soft Supersymmetry-Breaking Parameter RGEs From One-Loop 
Observable-Sector Contributions} 

%%%%%%%%%%%%%%%%%%%%%%%%%%%%%%%%%%%%%%%%%%%%%%%%%%%%%%%%
In this Appendix we present the one-loop MSSM RGEs including gauge-singlet Majorana 
neutrinos. Hidden-sector contributions to the RGE flow should be added to these (for the soft 
scalar mass-squared terms) following the discussion in Sections 3 and 4 of the text. 
\begin{equation}
\label{eq-X}
\frac{dX}{dt}=\frac{1}{16\pi^2}\dot{X}
\end{equation}
where $X$ denotes any of
$g_1$, $g_2$, $g_3$,
$\Yn$, $\Ye$, $\Yu$, $\Yd$,
$M_1$, $M_2$, $M_3$,
$\mhus$, $\mhds$,
$\mls$, $\mns$, $\mes$, $\mqs$, $\mus$, $\mds$,
$\An$, $\Ae$, $\Au$, $\Ad$. The dotted quantities appear below:
% %
% %
\begin{eqnarray}
\label{eq-g1}
\dot g_1 &=& 11 g_1^3, \\
\label{eq-g2}
\dot g_2 &=& g_2^3, \\
\label{eq-g3}
\dot g_3 &=& -3g_3^3, \\
\end{eqnarray}
% %
Yukawa couplings
% %
\begin{equation}
\label{eq-Yn}
\dot\Yn=\Yn\left(
- g_1^2 \unit
- 3 g_2^2 \unit
+ 3\tr\left(\Yu^\dagger \Yu \right) \unit
+ \tr\left(\Yn^\dagger \Yn \right) \unit
 + 3 \Yn^\dagger \Yn
 + \Ye^\dagger \Ye
 \right),
 \end{equation}
% %
% %
\begin{equation}
\label{eq-Ye}
\dot\Ye=\Ye\left(
- 3 g_1^2 \unit
- 3 g_2^2 \unit
+ 3 \tr\left(\Yd^\dagger \Yd \right) \unit
+ \tr\left(\Ye^\dagger \Ye \right) \unit
+ 3 \Ye^\dagger \Ye
+ \Yn^\dagger \Yn
\right),
\end{equation}
% %
% %
\begin{eqnarray}
\label{eq-Yu}
\dot\Yu &=& \Yu\left(
- \frac{13}{9} g_1^2 \unit
- 3 g_2^2 \unit
- \frac{16}{3} g_3^2 \unit
+ 3 \tr\left(\Yu^\dagger \Yu \right) \unit
+ \tr\left(\Yn^\dagger \Yn \right) \unit \right. \nonumber\\
& & + \left. 3 \Yu^\dagger \Yu
+ \Yd^\dagger \Yd
\right),
\end{eqnarray}
% %
% %
\begin{eqnarray}
\label{eq-Yd}
\dot\Yd &=& \Yd\left(
- \frac{7}{9} g_1^2 \unit
- 3 g_2^2 \unit
- \frac{16}{3} g_3^2\unit
+ 3 \tr\left(\Yd^\dagger \Yd \right) \unit
+ \tr\left(\Ye^\dagger \Ye \right) \unit \right. \nonumber \\
& & + \left. 3 \Yd^\dagger \Yd
+ \Yu^\dagger \Yu
\right),
\end{eqnarray}
% %
%% gaugino soft masses
% %
\begin{eqnarray}
\label{eq-M1}
\dot M_1 &=& 22 g_1^2 M_1, \\
\label{eq-M2}
\dot M_2 &=& 2 g_2^2 M_2, \\
\label{eq-M3}
\dot M_3 &=& -6 g_3^2 M_3,
\end{eqnarray}
% %
% %
\begin{equation}
\label{eq-S}
S=\mhus-\mhds+\tr\left(\mqs-2\mus+\mds-\mls+\mes\right),
\end{equation}
% %
up and down Higgs soft masses
% %
\begin{eqnarray}
\label{eq-mhu2}
\dot\mhus &=&
6 \tr\left(
  \mqs \Yu^\dagger \Yu
+ \Yu^\dagger \mus \Yu
+ \mhus \Yu^\dagger \Yu
+ \Au^\dagger \Au
\right) \nonumber \\
& & + 2 \tr\left(
  \mls \Yn^\dagger \Yn
+ \Yn^\dagger \mns \Yn
+ \mhus \Yn^\dagger \Yn
+ \An^\dagger \An
\right) \nonumber \\
& & -2g_1^2 M_1^2
- 6 g_2^2 M_2^2
+ g_1^2 S + {(2\lambda^2 \mhus)},
\end{eqnarray}
% %
\begin{eqnarray}
\label{eq-mhd2}
\dot\mhds &=&
2 \tr\left(
  \mls \Ye^\dagger \Ye
+ \Ye^\dagger \mes \Ye
+ \mhds \Ye^\dagger \Ye
+ \Ae^\dagger \Ae
\right) \nonumber \\
& & + 6 \tr\left(
\mqs \Yd^\dagger \Yd
+ \Yd^\dagger \mds \Yd
+ \mhds \Yd^\dagger \Yd
+ \Ad^\dagger \Ad
\right) \nonumber \\
& & - 2 g_1^2 M_1^2
- 6 g_2^2 M_2^2 - g_1^2 S + {(2\lambda^2 \mhds)},
\end{eqnarray}
% %
% % slepton soft masses
% %
\begin{eqnarray}
\label{eq-ml2}
\dot\mls &=&
\mls \Ye^\dagger \Ye
+ \Ye^\dagger \Ye \mls
+ \mls \Yn^\dagger \Yn
+ \Yn^\dagger \Yn \mls \nonumber \\
& & + 2 \Ye^\dagger \mes \Ye
+ 2 \mhds \Ye^\dagger \Ye
+ 2 \Ae^\dagger \Ae \nonumber \\
& & + 2 \Yn^\dagger \mns \Yn
+ 2 \mhus \Yn^\dagger \Yn
+ 2 \An^\dagger \An \nonumber \\
& & - 2 g_1^2 M_1^2 \unit
- 6 g_2^2 M_2^2 \unit
- g_1^2 S \unit + {(2\lambda^2 \mls)},
\end{eqnarray}
% %
\begin{equation}
\label{eq-mn2}
\dot\mns =
2 \mns \Yn \Yn^\dagger
+ 2 \Yn \Yn^\dagger \mns
+ 4 \Yn \mls \Yn^\dagger
+ 4 \mhus \Yn \Yn^\dagger
+ 4 \An \An^\dagger + {(2\lambda^2 \mns)},
\end{equation}
% %
% %
\begin{eqnarray}
\label{eq-me2}
\dot\mes &=&
2 \mes \Ye \Ye^\dagger
+ 2 \Ye \Ye^\dagger \mes
+ 4 \Ye \mls \Ye^\dagger
+ 4 \mhds \Ye \Ye^\dagger
+ 4 \Ae \Ae^\dagger \nonumber \\
& & - 8 g_1^2 M_1^2 \unit
+ 2 g_1^2 S \unit + {(2\lambda^2 \mes)},
\end{eqnarray}
% %
% % squark masses
% %
\begin{eqnarray}
\label{eq-mq2}
\dot\mqs &=&
\mqs \Yu^\dagger \Yu
+ \Yu^\dagger \Yu \mqs
+ 2 \Yu^\dagger \mus \Yu
+ 2 \mhus \Yu^\dagger \Yu
+ 2 \Au^\dagger \Au \nonumber \\
& & + \mqs \Yd^\dagger \Yd
+ \Yd^\dagger \Yd \mqs
+ 2 \Yd^\dagger \mds \Yd
+ 2 \mhds \Yd^\dagger \Yd
+ 2 \Ad^\dagger \Ad \nonumber \\
& & - \frac{2}{9} g_1^2 M_1^2 \unit
- 6 g_2^2 M_2^2 \unit
- \frac{32}{3} g_3^2 M_3^2 \unit
+ \frac{1}{3} g_1^2 S \unit + {(2\lambda^2 \mqs)},
\end{eqnarray}
% %
% %
\begin{eqnarray}
\label{eq-mu2}
\dot\mus &=&
2 \mus \Yu \Yu^\dagger
+ 2 \Yu \Yu^\dagger \mus
+ 4 \Yu \mqs \Yu^\dagger
+ 4 \mhus \Yu \Yu^\dagger
+ 4 \Au \Au^\dagger \nonumber \\
& & - \frac{32}{9} g_1^2 M_1^2 \unit
- \frac{32}{3} g_3^2 M_3^2 \unit
- \frac{4}{3} g_1^2 S \unit + {(2\lambda^2 \mus)},
\end{eqnarray}
% %
% %
\begin{eqnarray}
\label{eq-md2}
\dot\mds &=&
2 \mds \Yd \Yd^\dagger
+ 2 \Yd \Yd^\dagger \mds
+ 4 \Yd \mqs \Yd^\dagger
+ 4 \mhds \Yd \Yd^\dagger
+ 4 \Ad \Ad^\dagger \nonumber \\
& & - \frac{8}{9} g_1^2 M_1^2 \unit
- \frac{32}{3} g_3^2 M_3^2 \unit
+ \frac{2}{3} g_1^2 S \unit + {(2\lambda^2 \mds)},
\end{eqnarray}
% %
% % trilinear couplings
% %
\begin{eqnarray}
\label{eq-An}
\dot\An &=&
-g_1^2 \An
-3 g_2^2 \An
+3 \tr\left( \Yu^\dagger \Yu \right) \An
+ \tr\left( \Yn^\dagger \Yn \right) \An \nonumber \\
& & - 2 g_1^2 M_1 \Yn
- 6 g_2^2 M_2 \Yn
+ 6 \tr\left( \Yu^\dagger \Au \right) \Yn
+ 2 \tr\left( \Yn^\dagger \An \right) \Yn \nonumber \\
& & + 4 \Yn \Yn^\dagger \An
+ 5 \An \Yn^\dagger \Yn
+ 2 \Yn \Ye^\dagger \Ae
+ \An \Ye^\dagger \Ye,
\end{eqnarray}
% %
% %
\begin{eqnarray}
\label{eq-Ae}
\dot\Ae &=&
- 3 g_1^2 \Ae
- 3 g_2^2 \Ae
+ 3 \tr\left( \Yd^\dagger \Yd \right) \Ae
+ \tr\left( \Ye^\dagger \Ye \right) \Ae \nonumber\\
& & -6 g_1^2 M_1 \Ye
- 6 g_2^2 M_2 \Ye
+ 6 \tr\left( \Yd^\dagger \Ad \right) \Ye
+ 2 \tr\left( \Ye^\dagger \Ae \right) \Ye \nonumber\\
& & + 4 \Ye \Ye^\dagger \Ae
+ 5 \Ae \Ye^\dagger \Ye
+ 2 \Ye \Yn^\dagger \An
+ \Ae \Yn^\dagger \Yn,
\end{eqnarray}
% %
% %
\begin{eqnarray}
\label{eq-Au}
\dot\Au &=&
- \frac{13}{9} g_1^2 \Au
- 3 g_2^2 \Au
- \frac{16}{3} g_3^2 \Au
+ 3 \tr\left( \Yu^\dagger \Yu \right) \Au
+ \tr\left( \Yn^\dagger \Yn \right) \Au \nonumber\\
& & - \frac{26}{9} g_1^2 M_1 \Yu
- 6 g_2^2 M_2 \Yu
- \frac{32}{3} g_3^2 M_3 \Yu
+ 6 \tr\left( \Yu^\dagger \Au \right) \Yu
+ 2 \tr\left( \Yn^\dagger \An \right) \Yu \nonumber\\
& & + 4 \Yu \Yu^\dagger \Au
+ 5 \Au \Yu^\dagger \Yu
+ 2 \Yu \Yd^\dagger \Ad
+ \Au \Yd^\dagger \Yd,
\end{eqnarray}
% %
% %
\begin{eqnarray}
\label{eq-Ad}
\dot\Ad &=&
- \frac{7}{9} g_1^2 \Ad
- 3 g_2^2 \Ad
- \frac{16}{3} g_3^2 \Ad
+ 3 \tr\left( \Yd^\dagger \Yd \right) \Ad
+ \tr\left( \Ye^\dagger \Ye \right) \Ad \nonumber\\
& & - \frac{14}{9} g_1^2 M_1 \Yd
- 6 g_2^2 M_2 \Yd
- \frac{32}{3} g_3^2 M_3 \Yd
+ 6 \tr\left( \Yd^\dagger \Ad \right) \Yd
+ 2 \tr\left( \Ye^\dagger \Ae \right) \Yd \nonumber\\
& & + 4 \Yd \Yd^\dagger \Ad
+ 5 \Ad \Yd^\dagger \Yd
+ 2 \Yd \Yu^\dagger \Au
+ \Ad \Yu^\dagger \Yu.
\end{eqnarray}
% 
%%%%%%%%%%%%%%%%%%%%%%%%%%%%%%%%%%%%%%%%%%%%%%%%%%%%%%%
%%%%%%%%%%%%%%%%%%%%%%%%%%%%%%%%%%%%%%%%%%%%%%%%%%%%%%%
%%%%%%%%%%%%%%%%%%%%%%%%%%%%%%%%%%%%%%%%%%%%%%%%%%%%%%%

%%%%%%%%%%%%%%%%%%%%%%%%%%%%%%%%%%%%%%%%%%%%%%%%%%%%%%%
%%%%%%%%%%%%%%%%%%%%%%%%%%%%%%%%%%%%%%%%%%%%%%%%%%%%%%%

\end{document}